%% file: main.tex
\documentclass[camera,letterpaper,nomarginnotes,nonarrowgutter]{jpaper}
\usepackage{mathptmx} 
\newcommand{\ignore}[1]{}
\usepackage{fancyhdr}
\usepackage{cite}
\usepackage[normalem]{ulem}
\usepackage{siunitx} 
\usepackage[toc,page]{appendix}

\usepackage{lipsum}
\usepackage{makecell}
\newcommand{\stripe}{\rowcolor{blue!5}}

\usepackage{rotating}

\usepackage[bookmarks=true,breaklinks=true,letterpaper=true,colorlinks,linkcolor=black,citecolor=blue,urlcolor=black]{hyperref}

\usepackage[dvipsnames]{xcolor}
\usepackage{graphicx}
\usepackage[export]{adjustbox}
\usepackage{subfig}
\usepackage{afterpage}
\usepackage{placeins}

\usepackage{mathptmx} 
\usepackage{fancyhdr}
\usepackage[normalem]{ulem}
\usepackage{colortbl}
\usepackage{enumitem}
\usepackage{lipsum}
\usepackage{amsmath}
\usepackage{setspace}
\usepackage{cite}
\usepackage{soul}
\usepackage{tablefootnote}
\usepackage{threeparttable}
\usepackage{gensymb}
\usepackage[all]{nowidow}

\usepackage[export]{adjustbox}

\usepackage[compatibility=false, skip=0pt]{caption}
\usepackage{comment}
\usepackage{xspace}

\newcommand{\mechanism}{CODIC\xspace}
\newcommand{\mechanismLarge}{Fine-Grained Control Over DRAM Internal Circuit Timings}

\soulregister\cite7
\soulregister\ref7
\soulregister\pageref7

\colorlet{soulyellow}{yellow!60}
\colorlet{soulcyan}{cyan!30}
\colorlet{soulgreen}{green!30}

\colorlet{shadecolor}{yellow!60}
\colorlet{framecolor}{red}
    {\endMakeFramed}
\newenvironment{frshaded*}{%
    \MakeFramed {\advance\hsize-\width \FrameRestore}}%
    {\endMakeFramed}

\newlength\MyIndent
\newcommand{\upla}[1]{\mechanism{}-sig}
\newcommand{\uesa}[1]{\mechanism{}-sigsa}
\newcommand{\dtran}[1]{\mechanism{}-det}

\newcommand{\uesaLong}[1]{\underline{U}npredictable Values in the DRAM \underline{S}As}
\newcommand{\uplaLong}[1]{\underline{U}npredictable Values in the DRAM \underline{C}ells}
\newcommand{\dtranLong}[1]{\underline{D}eterministic Values}

\newcommand{\UPLAPUF}[1]{\mechanism{}-sig PUF}
\newcommand*\circled[1]{\tikz[baseline=(char.base)]{
            \node[shape=circle,draw,inner sep=0pt,fill=black, text=white] (char) {#1};}}
            
\makeatletter
\newcommand\footnoteref[1]{\protected@xdef\@thefnmark{\ref{#1}}\@footnotemark}
\makeatother
            
\newif\ifcameraready
\camerareadytrue

\newif\ifsubmission
\submissiontrue

\ifsubmission
\newcommand{\lois}[1]{\textcolor{black}{#1}}
\newcommand{\minp}[1]{\textcolor{black}{#1}} 
\definecolor{dollarbill}{rgb}{0.52,0.73,0.4}
\newcommand\nikaBox[1]{}
\newcommand{\todo}[1]{}

\definecolor{Moegi}{rgb}{0.357, 0.537, 0.188}

\newcommand{\onur}[1]{\textcolor{black}{#1}}
\newcommand{\om}[1]{\textcolor{black}{#1}}
\newcommand{\omm}[1]{\textcolor{black}{#1}}
\newcommand{\ommm}[1]{\textcolor{black}{#1}}
\newcommand{\ommmm}[1]{\textcolor{black}{#1}}
\newcommand{\ommmmm}[1]{\textcolor{black}{#1}}

\newcommand{\loiss}[1]{\textcolor{black}{#1}}
\newcommand{\loisss}[1]{\textcolor{black}{#1}}

\else
\definecolor{amber}{rgb}{1.0, 0.49, 0.0}
\newcommand{\lois}[1]{\textcolor{magenta}{#1}} 
\newcommand{\loiss}[1]{\textcolor{blue}{#1}} 
\definecolor{dollarbill}{rgb}{0.52,0.73,0.4}

\newcommand{\minp}[1]{\textcolor{yellow}{#1}} 
\newcommand\nikaBox[1]{\noindent{\color{blue} {\fbox{\bf{Nika: }}\it"#1"}}} 
\newcommand{\todo}[1]{\textbf{\textcolor{red}{[TODO] #1}}}
\definecolor{Moegi}{rgb}{0.357, 0.537, 0.188}

\newcommand{\onur}[1]{\textcolor{magenta}{#1}}
\newcommand{\loiss}[1]{\textcolor{black}{#1}}
\newcommand{\loisss}[1]{\textcolor{black}{#1}}

\fi

\expandafter\def\expandafter\UrlBreaks\expandafter{\UrlBreaks
  \do\a\do\b\do\c\do\d\do\e\do\f\do\g\do\h\do\i\do\j
  \do\k\do\l\do\m\do\n\do\o\do\p\do\q\do\r\do\s\do\t
  \do\u\do\v\do\w\do\x\do\y\do\z\do\A\do\B\do\C\do\D
  \do\E\do\F\do\G\do\H\do\I\do\J\do\K\do\L\do\M\do\N
  \do\O\do\P\do\Q\do\R\do\S\do\T\do\U\do\V\do\W\do\X
  \do\Y\do\Z}

\newcommand{\affilETH}[0]{\textsuperscript{\S}}
\newcommand{\affilNUDT}[0]{\textsuperscript{$\ddagger$}}
\newcommand{\affilIPM}[0]{\textsuperscript{$\dagger$}}

\newcommand{\affililli}[0]{\textsuperscript{$\mp$}}

 \author{
 {Lois Orosa\affilETH}\qquad%
 {Yaohua \om{Wang\affilNUDT}}\qquad%
 {Mohammad Sadrosadati\affilIPM\affilETH}\qquad%
 {Jeremie \onur{S.} Kim\om{\affilETH}}\\%
 {Minesh Patel\affilETH}\qquad%
 {Ivan Puddu\affilETH}\qquad%
 {Haocong Luo\affilETH}\qquad%
 {Kaveh Razavi\affilETH}\qquad%
 {Juan\ G\'omez-Luna\affilETH}\\%
 {Hasan Hassan\affilETH}\qquad%
 {Nika Mansouri-Ghiasi\affilETH}\qquad%
 {Saugata Ghose\affililli}\qquad%
 {Onur Mutlu\om{\affilETH}}\vspace{2pt}\\%
 {\small\it\affilETH ETH Z{\"u}rich  \qquad \omm{\affilNUDT National University of Defense Technology (NUDT)}   }
 \\%
 {\small\it\omm{\affilIPM Institute for Research in Fundamental Sciences (IPM) \qquad \affililli University of Illinois at Urbana--Champaign}}
 \vspace{-1em}%
 }

\title{\vspace{-1.5em}\mechanism{}: A Low-Cost Substrate for Enabling\\ Custom \lois{In}-DRAM  Functionalities and Optimizations}

\begin{document}

\setstretch{0.89}

\cfoot{\thepage}
\sloppy

\maketitle

\definecolor{MidnightBlue}{rgb}{0.1, 0.1, 0.44}
\newcommand{\versionnum}[0]{4.4~---~\today~@~\currenttime~CET} 
\fancyhead{}
\ifcameraready
 \thispagestyle{plain}
 \pagestyle{plain}
\else
 \fancyhead[C]{\textcolor{MidnightBlue}{\emph{ISCA 2021 arXiv Camera Ready Version \versionnum{}}}}
 \fancypagestyle{firststyle}
 {
   \fancyhead[C]{\textcolor{MidnightBlue}{\emph{ISCA 2021 arXiv Camera Ready Version \versionnum{}}}}
   \fancyfoot[C]{\thepage}
 }
 \thispagestyle{firststyle}
 \pagestyle{firststyle}
\fi



\setstretch{0.90}
\input{00_abstract}

\input{01_introduction}
\input{02_background}
\input{03_motivation}

\setstretch{0.89}
\input{04_substrate}
\setstretch{0.887}
\input{05_applications}

\input{06_evaluation}
\input{08_related}
\input{09_conclusion}

\section*{Acknowledgments} 
\om{We thank the anonymous reviewers of \ommmm{ISCA\textquotesingle21/19/18, HPCA\textquotesingle21/19/18, MICRO\textquotesingle20/18, IEEE S\&P\textquotesingle20, and USENIX Security\textquotesingle19} for feedback, and the} SAFARI Research Group members for {valuable} feedback and the stimulating intellectual environment they provide. \onur{We thank Uksong Kang, Vivek Seshadri, and Arash Tavakkol for their feedback and suggestions on earlier versions of this paper}. We acknowledge the generous gifts provided by our industrial partners: Google, Huawei, Intel, Microsoft, and VMware. \onur{A \om{much earlier version of this paper \omm{was placed on}} arXiv in February 2019~\cite{orosa2019dataplant}.}

\bibliographystyle{IEEEtranS}
\bibliography{main}

\renewcommand*\appendixpagename{Appendix}
\begin{appendices}
\input{10_secureDeallocation}
\input{12_NIST}

\input{13_newCODIC_variant}
\input{11_DRAMchips}

\end{appendices}

\end{document}

%% file: 00_abstract.tex
\begin{abstract}

\onur{DRAM is the dominant main memory technology used \om{in} modern computing systems. 
Computing systems implement a memory controller that interfaces with DRAM via DRAM commands. DRAM executes the given commands using internal components (e.g., access \om{transistors}, sense \om{amplifiers}) that are orchestrated by DRAM internal timings, which are fixed for each DRAM command.
Unfortunately, the use of fixed internal timings limits the types of operations that DRAM can perform and hinders the implementation of new functionalities and custom mechanisms that improve DRAM reliability, performance and energy. To overcome these limitations, we propose enabling programmable DRAM internal timings for controlling in-DRAM components.}

To this end, we design \mechanism{}, a new low-cost DRAM substrate that enables fine-grained control over four previously fixed internal DRAM timings that are key to many DRAM operations. We implement \mechanism{} with only minimal changes to the DRAM \om{chip} and the DDRx interface.
To demonstrate the potential of \mechanism{}, we \onur{propose two new \mechanism{}-based} security mechanisms that outperform state-of-the-art mechanisms in several ways: (1)~a new DRAM Physical Unclonable Function (PUF) that is more \om{robust} and has significantly higher throughput than state-of-the-art DRAM PUFs, and (2)~the first cold boot attack prevention mechanism that does not introduce any \onur{performance or energy overheads} at runtime. 

\end{abstract}

%% file: 01_introduction.tex
\section{Introduction}

DRAM is the ubiquitous technology employed for main memory across computing systems. System components interact with DRAM via DRAM commands. These commands trigger a set of internal DRAM circuit signals that control different DRAM components in a timely manner to achieve a specific functionality (e.g., activate a DRAM row). The timing with which these signals are triggered and the relative order of these signals determine the functionality, energy consumption, and latency of the resulting operation. On modern DRAM devices, the manufacturer fixes these timings at design time \lois{for each command}. However, having fixed internal circuit timings for each DRAM operation limits the operations that \onur{the internal DRAM circuits} can perform and hinders \lois{their} potential for \lois{implementing} custom optimizations \lois{that adapt} to the characteristics of the device (e.g., process variation) and environmental conditions (e.g., temperature changes, aging).

The lack of control over internal DRAM circuit timings inhibits \lois{at least} two research directions. 
First, prior works that extend the \om{functionalities of DRAM~\cite{seshadri2017ambit, seshadri2013, gao2019computedram,seshadri2015fast,kim2019d,kim2018,hassan2019crow,choi2015multiple,Hassan2016,seshadri2016buddy}} cannot be evaluated or demonstrated reliably on real DRAM chips due to the lack of control over the DRAM \omm{internal} circuit timing~\cite{keeth2008dram}. 
One example of this limitation is ComputeDRAM~\cite{gao2019computedram}, which demonstrates computing capabilities in DRAM by triggering a specific sequence of commands from the memory controller with modified timing parameters. Although \onur{ComputeDRAM} shows the potential for extending functionality by varying standard DRAM operating timings, 
\onur{the results shown in the paper demonstrate that only a small fraction of the cells can reliably perform the intended computations; for a vast majority of cells the intended computation cannot be performed reliably.}
The reason for this unreliability remains unknown, in part because of the black-box design of DRAM that completely limits the user's visibility \onur{into} and control over the \omm{DRAM internal} timing signals. 

Second, several works propose DRAM energy and access latency optimizations by reducing timing parameters \emph{of commands issued by the memory controller}\om{~\cite{Lee2015,Hassan2016,Chang:2016,zhang2016restore,Lee:2017,wang2018reducing, kim2018solar,koppula2019eden}}. However, these works are limited to \lois{modifications to memory controller timing parameters}, and they do not \lois{explore optimizations of internal in-DRAM timing signals}. For example, an activate command might have a very conservative time interval between \onur{(1)~the internal in-DRAM signals that activate the access transistors of  DRAM cells, and (2)~the internal in-DRAM signals that activate the sense amplifiers (SA)} for sensing and amplifying the content of DRAM cells. \lois{Reducing this time interval for the DRAM regions that can reliably operate with reduced in-DRAM timings can \om{further improve} overall system performance.}

In this work, we advocate for a substrate that enables greater control over DRAM internal circuit timings as an efficient and low-cost way to enable new functionalities and optimizations in DRAM. 
We propose \om{\mechanism{},\footnote{\lois{Fine-grained \ul{CO}ntrol over \ul{D}RAM \ul{I}nternal \ul{C}ircuit timings \om{(CODIC).}}}} a low-cost DRAM substrate that enables fine-grained control of four key signals that \onur{orchestrate} DRAM internal circuit timings. 
The circuits controlled by \mechanism{} perform fundamental operations that \onur{(1)}~connect DRAM cells to bitlines, \onur{(2)}~trigger sense amplifiers, and \onur{(3)}~trigger the logic to prepare DRAM banks for the next access. 
\lois{\onur{The memory controller} can access the substrate with a new \mechanism{} command, which can \onur{be} configured with in-DRAM registers to enable new functionalities and optimizations.}

By providing \onur{fine-granularity} control over DRAM \lois{internal circuit timings}, \mechanism{} enables new DRAM functionalities and optimizations of existing DRAM commands. To demonstrate this point, we implement and evaluate in detail two \onur{configurations} of the \mechanism{} \onur{substrate} that expose two new functionalities: 
\ommmmm{{(1)~in-place generation of digital signatures, which depend on process variation and are unique to each DRAM region and DRAM device (i.e., Physical Unclonable Functions, PUFs~\cite{Gassend:2002,Daihyun2005,Suh2007,kim2018}}); and (2)~in-place generation of deterministic values, which are defined at DRAM design-time.} \lois{These two functionalities are key for implementing several security applications.}

\lois{Although \mechanism{} can be used for \lois{multiple} purposes, such as reducing DRAM latency and energy or improving existing DRAM mechanisms (see Section~\ref{sec:otheraplications}), \lois{we find that} \mechanism{} \lois{can be very useful for improving} the security of two specific emerging computing platforms.}
First, low-power systems (e.g., Internet of Things or IoT devices) cannot afford sophisticated security mechanisms adopted by high-end processors (e.g., memory  \om{encryption~\cite{costan2016intel}}) because such mechanisms can consume significant energy and \ommmm{area~\cite{Zhang2014}}. Second, \onur{Processing-In-Memory (PIM)} systems \ommmm{(e.g.,~\cite{ahn2015scalable,ahn2015pim,li2016pinatubo,seshadri2017ambit,seshadri2013,mutlu2020modern,ghose2019processing,seshadri2019dram,mutlu2019processing,ghose2018enabling,seshadri2017simple,hajinazar2021simdram,seshadri2016buddy,seshadri2015fast,oliveira2021damov,juan2021benchmarking,boroumand2019conda,boroumand2018google,boroumand2021mitigating,fromm1997energy,stone1970logic,kang1999flexram,gokhale1995processing})} implement low-cost computing units that typically do not have security features and lack an interface to access the security features of the host processor. For both types of systems, \mechanism{} enables implementing security support \emph{within DRAM}, thereby enabling suitable applications for improving system security at low-cost.

\lois{As a proof of concept,} we use \mechanism{} to develop two \onur{new} applications for improving system security. First, we propose a new \mechanism{}-based Physical Unclonable Function \om{(PUF)~\cite{Gassend:2002,Daihyun2005,Suh2007,kim2018}} with greater throughput and reliability than state-of-the-art DRAM PUFs\onur{~\cite{kim2018,talukder2019prelatpuf}}.
A PUF generates signatures unique to a device due to the unique physical variations of the device (i.e., process variation). PUFs are typically used to authenticate or uniquely identify a device. Our evaluation shows that our \mechanism{}-based PUF provides higher throughput, excellent resilience to temperature changes, and more \lois{stable} PUF responses \lois{(i.e., the same challenge produces the same response)} than the state-of-the-art DRAM PUFs\onur{~\cite{kim2018,talukder2019prelatpuf}}. 

Second, we propose a new \mechanism{}-based mechanism to prevent Cold Boot \om{Attacks~\cite{Yitbarek2017,Halderman2009,Simmons2011,Gruhn2013,BAUER2016S65,muller2010aesse,villanueva2019cold,mcgregor2008braving,lindenlauf2015cold,lee2011correcting,Hilgers2014}}. 
In a Cold Boot Attack, the attacker physically removes the DRAM module from the victim system and places it in a system under their control to extract secret information. Because data in DRAM is stored in capacitors, the data can potentially remain in the cells long enough (during physical extraction) for data to be stolen. The key idea of our \mechanism{}-based mechanism is to automatically overwrite the entire DRAM with \lois{values generated by \mechanism{}} when the DRAM chip is first \ommmm{powered-on}. \om{Our evaluation shows that our \mechanism{}-based mechanism  is \omm{2.0$\times$} faster than the best prior mechanism.}

The potential of \mechanism{} extends beyond the security applications proposed and evaluated in this work. \mechanism{} provides a low-cost substrate for controlling DRAM internal circuit timings that enables the implementation of a \onur{wide} variety of new functionalities, and \lois{custom optimizations \onur{that improve} reliability, performance, and energy consumption}. Section~\ref{sec:other_apps} discusses several \om{other} potential \mechanism{} use cases.

We make the following \om{major} contributions:

\begin{itemize}[leftmargin=3mm,itemsep=0mm,parsep=0mm,topsep=0mm]
\item We propose \mechanism{}, a new low-cost DRAM substrate for controlling DRAM internal circuit timings. \mechanism{} can be used to implement new functionalities, and custom optimizations \onur{that improve} reliability, performance, and energy consumption. To demonstrate the \lois{potential} of \mechanism{}, we propose and evaluate two \onur{configurations} of the \mechanism{} \onur{mechanism} that enable new security applications.

\item We propose a new \mechanism{}-based DRAM PUF that \onur{has} \om{1.8$\times$} \onur{higher throughput} than the best state-of-the-art DRAM PUF, has similar resilience to temperature changes, and provides more repeatable PUF responses than the state-of-the-art \om{DRAM PUF}. We evaluate our \mechanism{}-based DRAM PUF using 136 real commodity DRAM chips to \onur{validate} the \onur{functionality} of our mechanism.

\item We propose a new \mechanism{}-based cold boot attack prevention mechanism that operates only at \ommmm{power-on} while ensuring reliable protection. Our mechanism does not incur any latency or power overhead at runtime, 
and it \onur{is} \om{2.0$\times$} \onur{lower latency} and \om{1.7$\times$} \onur{lower energy than} the best state-of-the-art \om{mechanisms}\onur{~\cite{Chang2016LISA,seshadri2013}} during \ommm{DRAM power-on}.

\end{itemize}

%% file: 02_background.tex
\section{Background}
\label{sec:background}

\label{sec:dramorganization} 

\vspace{3pt}\noindent\textbf{DRAM Organization.} A single DRAM chip has limited capacity (e.g., 8Gb) and data width (e.g., 8-bit). DRAM chips are grouped together in a DRAM module to form a \emph{rank}, providing \onur{higher} capacity (e.g., 8GB) and larger data \onur{width} (e.g., \onur{64-bit}).

Each DRAM rank consists of multiple logical \emph{banks} striped across chips. \onur{Each} bank contains \onur{multiple (\om{typically 128}
to 512)} 2D DRAM cell arrays (i.e., \emph{subarrays}\om{~\cite{Kim:2012,Chang2014}}). 
\onur{Cells are organized in rows, which are connected to a row buffer (RB). The RB consists of a set of sense amplifiers that are used to activate one row at a time in the subarray. Each vertical line of cells is connected to one sense amplifier via a bitline wire. Cells within a row share a wordline. }
Each cell consists of a \emph{capacitor} \onur{that} stores data in form of charge ($V_{dd}$ or $0$V), and an \emph{access transistor} controlled by the wordline that connects \om{the cell} to the \emph{Sense Amplifier} through the \emph{bitline}.


\setlength\parfillskip{0pt plus .75\textwidth}
\setlength\emergencystretch{1pt}
\vspace{3pt}\noindent\textbf{DRAM Sense Amplifier (SA).} The Sense Amplifier (SA) is used to sense and amplify the small amount of charge in a DRAM cell capacitor to a CMOS-readable value. A set of SAs connected to a row of cells is called \emph{row buffer}. A cell is connected to \onur{an} SA via a \emph{bitline}. In the idle state, bitlines are driven to $V_{dd}$/2. When the SA is enabled, it detects any voltage \onur{deviation} from $V_{dd}$/2 and amplifies the \onur{bitline charge to} 0 or $V_{dd}$.

\begin{figure}[h] \centering
    \includegraphics[width=1.0\linewidth]{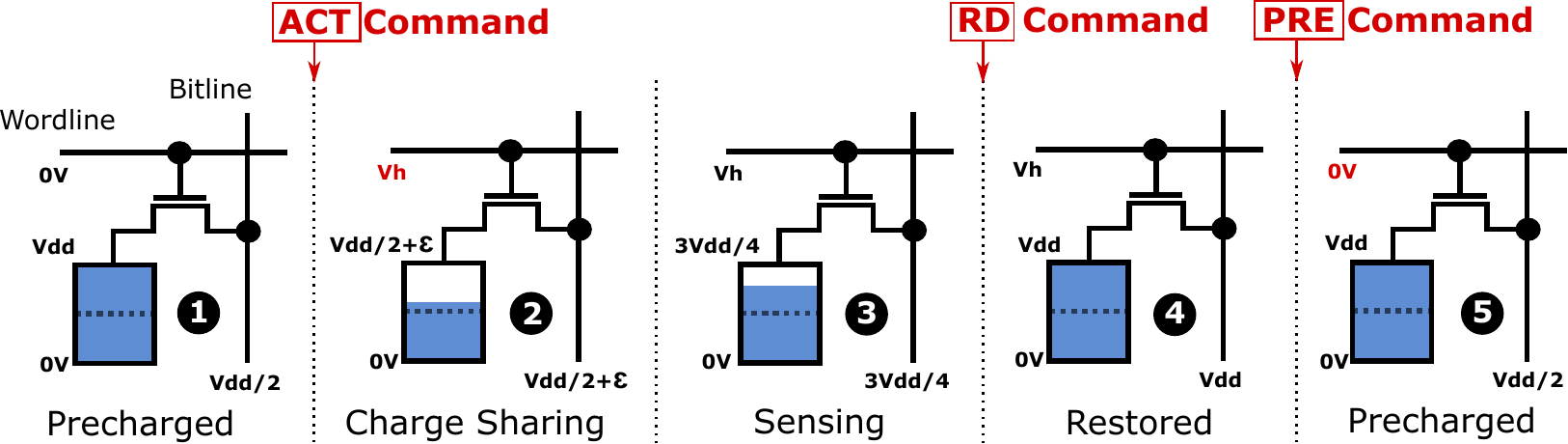}%
    \vspace{-0.1cm}%
    \caption{\om{Activate (ACT), Read (RD), Precharge (PRE) commands.}}%
    \label{fig:activation}
\end{figure}

\vspace{3pt}\noindent\textbf{DRAM Operation.} The memory controller \onur{interacts} with DRAM using three basic \onur{commands:} Activate (ACT), Read/Write (RD/WR), Precharge (PRE). Figure~\ref{fig:activation} details the steps for reading a DRAM cell using these commands. \circled{1} Initially, the bitline is held at $V_{dd}$/2 with the wordline at 0V. \circled{2} To access data, the memory controller issues an ACT command, which \onur{applies high voltage to} the target wordline, connecting that row's cells to their respective bitlines. This causes a deviation of the bitline voltage in one direction \lois{(i.e., \om{\emph{$\epsilon$}} in Figure~\ref{fig:activation}, \circled{2})}. \lois{This process is called \emph{charge sharing}}. \circled{3} Shortly after, the sense amplifier is enabled to sense and amplify this deviation \lois{(3$V_{dd}$/4 voltage represents an example of a voltage that is not fully amplified)}. After sufficient amplification, the memory controller can issue RD or WR commands \om{to the row buffer (i.e., sense amplifiers)}. The time needed to finish the ACT command is specified by the timing parameter $tRCD$. \circled{4} The sense amplifier continues to amplify the deviation until the voltage of the cell is fully restored \lois{(i.e., $V_{dd}$)}. \circled{5} Finally, the memory controller issues a PRE command to lower the wordline voltage back to 0V and drive the sense amplifier and bitline to $V_{dd}$/2. The time needed to complete a PRE command is specified by the timing parameter $tRP$. Once precharged, another row in the subarray can be \onur{activated}.

\begin{figure}[t]%
    \centering%
    \subfloat[Detailed SA \onur{circuit}]{
        \includegraphics[width=0.37\linewidth]{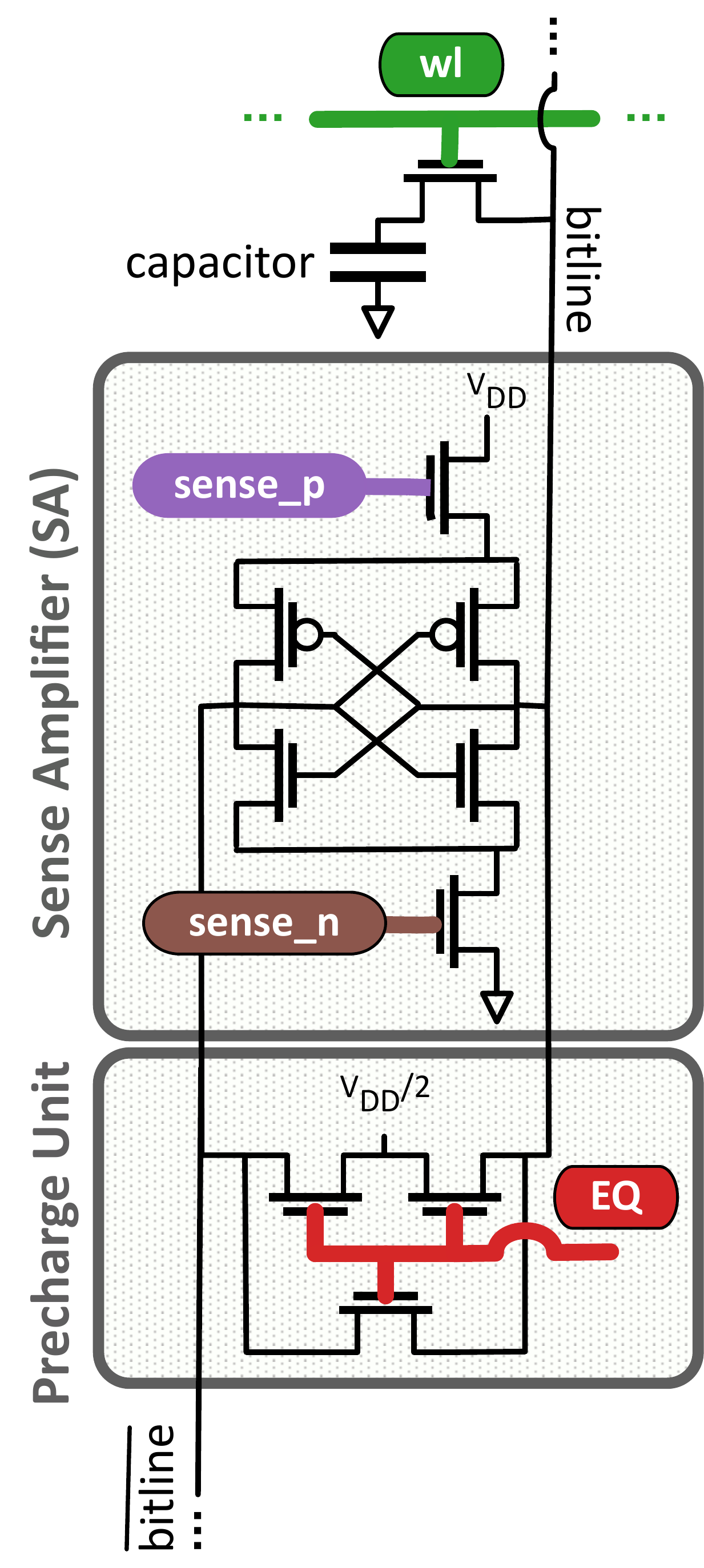}%
        \vspace{-5pt}%
    }%
    \subfloat[Precharge and activate commands]{
        {%
        \includegraphics[width=0.63\linewidth]{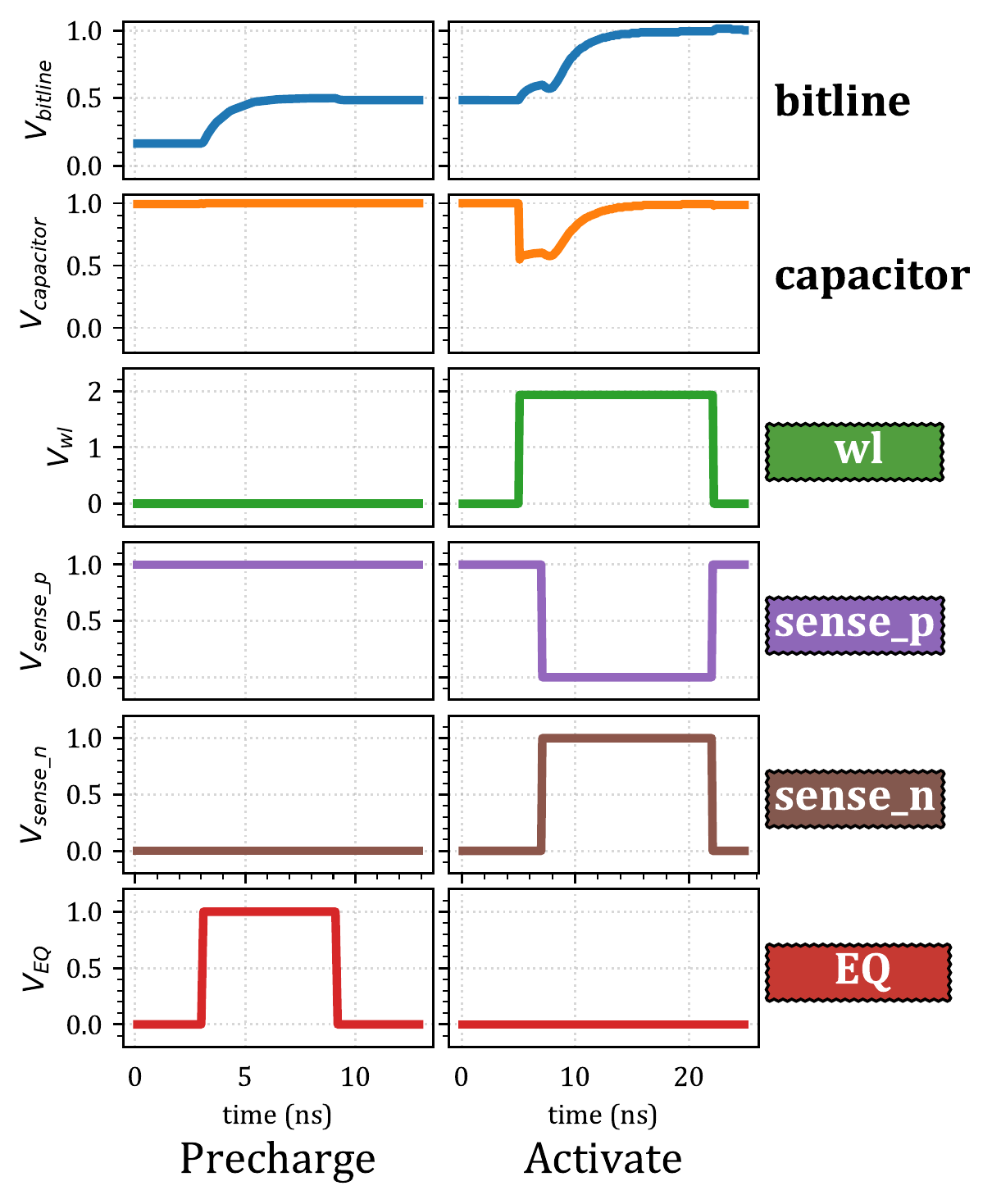}
        }
    }%
        \vspace{-0.1cm}
    \caption{(a) \om{Four fundamental} DRAM internal signals used to control the SA and precharge units, and (b) DRAM internal signal timing in regular precharge and activate commands.}%
    \label{fig:dram_signals}%
\end{figure}

\vspace{3pt}\noindent\textbf{DRAM Internal Circuit Timings.}
\om{Figure~\ref{fig:dram_signals}a} shows a detailed circuit of a common SA \om{design~\cite{keeth2008dram}}, and the four internal DRAM signals required to perform the activate and precharge commands: (1)~\emph{wl} controls the access transistor that connects the cell capacitor to the bitline, (2)~\emph{EQ} controls the precharge unit that sets the bitline to $V_{DD}/2$, (3)~\emph{sense\_p} controls the PMOS amplifier in the SA, and (4)~\emph{sense\_n} controls the NMOS amplifier in the SA.

\om{Figure~\ref{fig:dram_signals}b} shows the SPICE simulation of the four DRAM internal signals used to implement the precharge and activate commands, and their effect on the voltage of the bitline and the cell capacitor. During a PRE command, we observe that only \lois{the} EQ signal is triggered for setting the bitline to the precharge voltage level ($V_{dd}/2$). During the ACT command, we observe that (1)~the wordline \onur{(wl)} is triggered to share the charge between the cell and the bitline; and (2)~after charge sharing, sense\_n and sense\_p are triggered to sense and amplify \onur{the} small charge in the cell.

%% file: 03_motivation.tex
\section{Motivation and Goal}
\label{sec:motivation}

Many recent works \onur{reveal} via DRAM characterization that exploiting the effects of changing timings between DRAM commands can provide substantial benefits to the overall \mbox{system}~\cite{Lee2015,Hassan2016,Chang:2016,zhang2016restore,Lee:2017,wang2018reducing, kim2018solar,koppula2019eden}. However, prior works do not discuss the potential of controlling DRAM internal circuit timings.

\vspace{5pt}\noindent\textbf{\lois{Violating Memory Controller} DDRx Timing Parameters.}
We identify and describe three ways in which prior works exploit the ability to violate \lois{memory controller} DDRx timing parameters~\cite{jedec2012jedec}.
First, prior works~\cite{Lee2015,Hassan2016,Chang:2016,zhang2016restore,Lee:2017,wang2018reducing, kim2018solar,koppula2019eden} \om{identify} substantial margins in \lois{memory controller} DDRx timing parameters (i.e., DRAM latency \onur{settings}) that can be reduced to improve overall system performance. For example, Chang et al.~\cite{Chang:2016} experimentally demonstrate that they can statically profile DRAM and reduce \emph{tRCD} while maintaining data correctness.
Second, prior works~\cite{kim2018, talukder2019prelatpuf, kim2019d,olgun2021quactrng} exploit the characteristics of DRAM failures induced by accesses with reduced \lois{memory controller} DDRx timing parameters, to implement security features \onur{such as True Random Number Generators (TRNGs) and Physical Unclonable Functions (PUFs)}.
Third, Chang et al.~\cite{Chang:2017} experimentally demonstrate that they can improve the reliability of DRAM while operating at low voltages by increasing the \lois{memory controller} DDRx timing parameters.

\vspace{3pt}\noindent\textbf{Limitations of Fixed DRAM Internal Circuit Timings.}
While characterization of real devices in prior works has lead to important insights, these works are largely limited by the amount of control that the memory controller has over DRAM internal circuit timings.
We identify and describe two \lois{limitations} of current DRAM designs and the DDRx interface.

First, DRAM internal circuit timings (see Section~\ref{sec:background}) are chosen at design time and cannot be modified. DRAM manufacturers use very conservative \lois{internal circuit timings for implementing commands within DRAM (e.g., activate command)} with the goal of \lois{ensuring reliable operation} \onur{under} worst case conditions. Fixed DRAM internal circuit timings (1)~limit the ability to potentially further optimize DRAM commands when accounting for environmental conditions (e.g., temperature) or process variation within a DRAM device to improve DRAM latency (\om{e.g.,~\cite{Lee2015,Hassan2016,Chang:2016,zhang2016restore,Lee:2017,wang2018reducing, kim2018solar,koppula2019eden}}), and (2)~prevent the modification of these internal circuit timings, which can \om{otherwise} enable new functionalities~\cite{seshadri2017ambit, seshadri2013, gao2019computedram,seshadri2015fast,kim2019d,kim2018,hassan2019crow,choi2015multiple,Hassan2016,seshadri2016buddy} and research directions, as we demonstrate in this paper \lois{(see Section~\ref{sec:otheraplications})}.

\onur{Second, memory controller DDRx timing parameters can only control the time between issuing two DRAM commands. Although the memory controller can issue two DRAM commands with accurate timings, it does not have any knowledge or control over the internal implementation of DRAM commands, which limits the control and understanding of the effects of reducing DDRx timing parameters. For example, if an ACT command is issued with reduced DDRx timing parameters after \om{a} PRE command, the memory controller does not know the internal DRAM state of the on-going PRE command, which might have not finish its execution in DRAM when the ACT command is issued.}

\vspace{3pt}\noindent\textbf{The Potential of Controlling DRAM Internal Circuit Timings}.
Manufacturers provide a \onur{rigid} design without any type of control over the DRAM internal circuit timings. We advocate that having more control over the DRAM internal circuit timings has a lot of potential for enabling more aggressive performance, reliability, and energy optimizations, new functionalities, and may open new areas of research.

\vspace{3pt}\noindent\textbf{Goal.} Our goal in this paper is two-fold. First, we aim to provide a low-cost substrate in DRAM (\onur{called} \mechanism{}) that enables fine-grained control over DRAM internal signal timings, for enabling \onur{new} and enhancing existing DRAM commands and optimizations \lois{(Section~\ref{sec:substrate})}. Second, \lois{as a proof of concept}, we design \lois{two} new security mechanisms \lois{using \mechanism{}} that provide stronger security guarantees and higher performance \onur{over} state-of-the-art mechanisms \lois{(Section~\ref{sec:otheraplications})}. 

%% file: 04_substrate.tex
\section{\mechanism{} Substrate}
\label{sec:substrate}

We propose \mechanismLarge{} \onur{(\mechanism{})}, a new DRAM substrate that enables the control of four fundamental DRAM internal signals \om{that} control specific circuit timings that are otherwise fixed in DRAM. \mechanism{} allows \onur{fine-grained} control of these signals, which\om{, in turn,} enables the implementation of new DRAM functionalities and performance optimizations. 

The applications of \mechanism{} are \onur{general and many} (see Section~\ref{sec:otheraplications}). 
As a proof of concept, we demonstrate and evaluate two new variants \loiss{of \mechanism{}} that 
\lois{generate (1)~deterministic values that are defined at DRAM design-time, or (2)~unique digital signatures that depend on process variation.}
\lois{We use \mechanism{} to build} two new security applications that improve the state-of-the-art, as discussed in Sections~\ref{sec:physicallyUnclonable} and~\ref{sec:coldbootattacks}.

\subsection{Implementing New \mechanism{} Variants}
\label{sec:substrate:implementing}

\mechanism{} can  control four fundamental signals that control key internal circuit timings (\emph{wl}, \emph{EQ}, \emph{sense\_p}, and \emph{sense\_n}), which are described in \onur{Section~\ref{sec:background}}. \mechanism{} can trigger and disable these signals within a time window \lois{of 25ns}, \onur{at time steps} of 1ns \onur{(i.e., a signal can be triggered or disabled at [0ns, 1ns, ... 24ns] within the time window)}, which provides flexibility for implementing a wide variety of new commands, features, and optimizations.

Table~\ref{table:new_commands} shows two existing DRAM commands and two new \mechanism{} variants and the timings of the four signals. The third and fourth columns of the table show the signals that are triggered by the command, and the time (in ns) and direction ($\uparrow\downarrow$) that each signal is toggled. The two existing commands are the activate and precharge commands~\cite{jedec2012jedec}, and the two new \mechanism{} variants are \upla{}, for generating \lois{digital signatures that depend on process variation}, and \dtran{}, for generating deterministic values in-memory.

\begin{table}[h]
    \centering%
    \setlength\tabcolsep{5pt}
    \caption{In-DRAM signals used to implement activation and precharge~\cite{jedec2012jedec}, and two new \loiss{\mechanism{} variants}.}
    \vspace{-0.1cm}
    \scriptsize{}
    \begin{threeparttable}
    \begin{tabular}{ r  l | l  l }

        \toprule
        {\bf Command} & {\bf Description} &\multicolumn{2}{c}{\bf \om{Signal [init$\uparrow\downarrow$, end$\uparrow\downarrow$]}}\\
        \midrule
        \makecell[r]{Activation} & \loiss{See} Section~\ref{sec:dramorganization} & \om{wl [5$\uparrow$,22$\downarrow$]} & \makecell[l]{sense\_p\:[7$\downarrow$,22$\uparrow$]\\
        sense\_n~[7$\uparrow$,22$\downarrow$]} \\
        \midrule
        Precharge & \loiss{See} Section~\ref{sec:dramorganization} & \om{EQ [5$\uparrow$,11$\downarrow$]} &   \\
        \midrule
        {\bf \upla{}} & \makecell[l]{\loiss{Generates signature} values} & \om{wl [5$\uparrow$,22$\downarrow$]} & \om{EQ [7$\uparrow$,22$\downarrow$]}  \\
        \midrule
        {\bf \dtran{}} & \makecell[l]{\loiss{Generates} deterministic values} & \om{wl [5$\uparrow$,22$\downarrow$]} & \makecell[l]{sense\_p\:[14$\downarrow$,22$\uparrow$]\\ sense\_n~[7$\uparrow$,22$\downarrow$]}   \\
        \bottomrule
    \end{tabular}
    \end{threeparttable}
    \label{table:new_commands}
\end{table}

\subsubsection{\upla{}}
\label{sec:u-pla}

\lois{\mechanism{}-sig generates signature values that depend on process variation} by sensing and amplifying a DRAM cell that we set to the precharge voltage ($V_{dd}/2$). Sense amplifiers detect minor voltage differences above or below $V_{dd}/2$.
For a cell precharged to $V_{dd}/2$, the bit value that the sense amplifier converges to during activation is dependent on uncontrollable, \onur{truly-random} manufacturing process variation that perturbs the sensed voltage on the bitline. We verify this randomness in Section~\ref{sec:nist}, by showing that \lois{\mechanism{}-sig} values generated from \om{136} real DRAM chips fully pass the NIST randomness tests~\cite{rukhin2001statistical}.

\begin{figure}[h] \centering
    \includegraphics[width=1.0\linewidth]{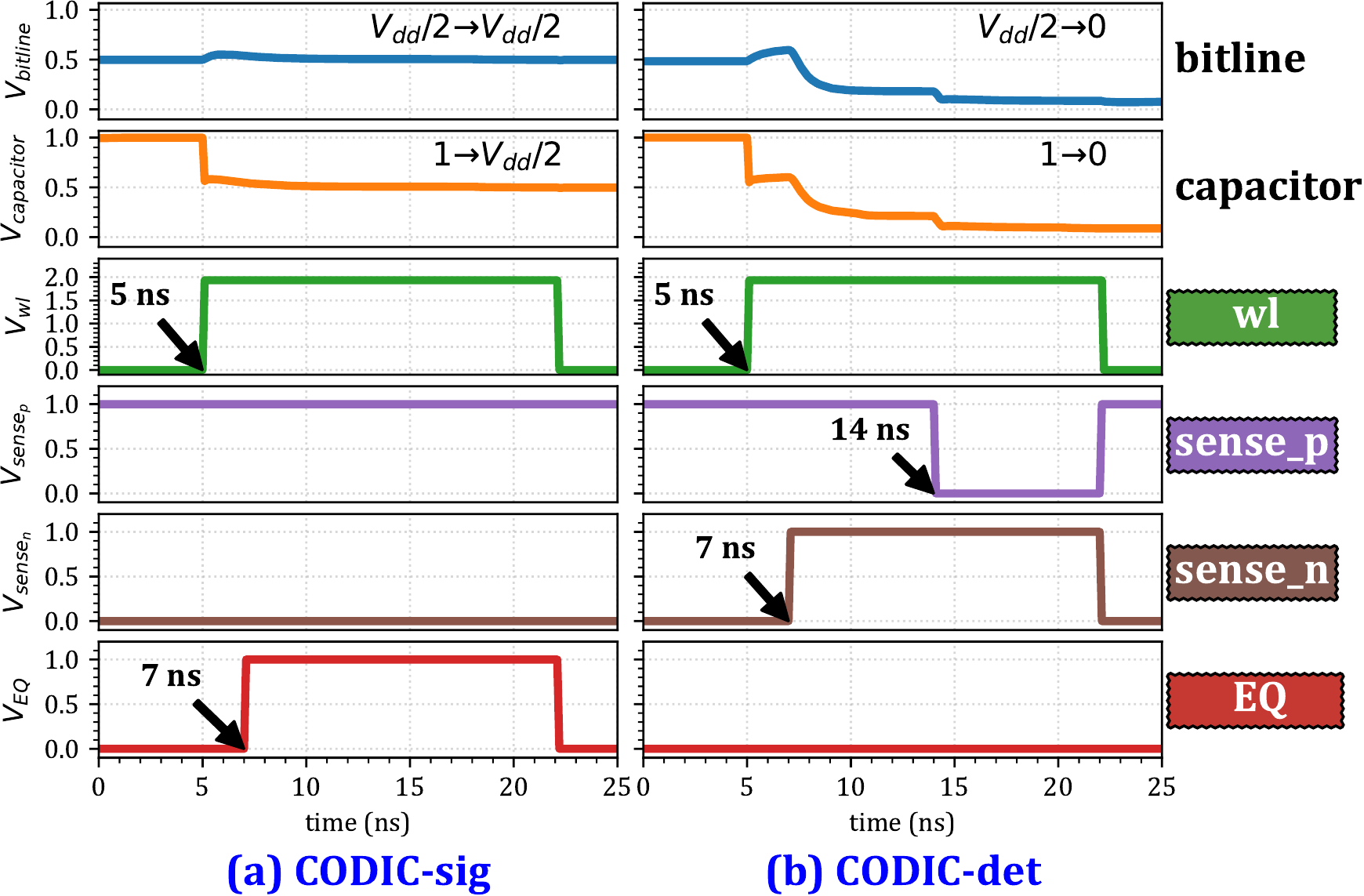}%
    \caption{Generating values in DRAM using \onur{two} \mechanism{} variants.}
    \label{fig:fleco_commands}
\end{figure}

Figure~\ref{fig:fleco_commands}\om{a} shows the SPICE simulation of \upla{}. \upla{} \onur{raises} the \emph{wl} signal (at 5 ns) before it raises the \emph{EQ} signal (at 7 ns), which drives the cell capacitor towards $V_{dd}$/2. Because the bitline is initially precharged to $V_{dd}$/2, it remains at $V_{dd}$/2 during the operation \om{(i.e., the bitline is in \omm{precharged} state)}. 
\onur{Only after the next activation command (not shown in Figure~\ref{fig:fleco_commands}\om{a}) the DRAM cell will be amplified to zero or one depending on process variation.}
We also propose {\bf \upla{}-opt}, a \upla{} optimization that is based on the key observation that \upla{} can set the voltage of the DRAM capacitor to $V_{dd}/2$ very quickly. Figure~\ref{fig:fleco_commands}\om{a} shows that the voltage of the capacitor is set to $V_{dd}$/2 almost immediately after \upla{} triggers the \emph{EQ} signal. Based on this observation, we can terminate the \emph{wl} and \emph{EQ} signals much earlier to reduce the latency of the command, without sacrificing reliability. \onur{To take} advantage of this optimization, the memory controller requires the ability to change the \mechanism{} timing parameters at runtime  (Section~\ref{sec:implementaion_details}).

%

\lois{We choose some specific timing values at which the \om{relevant signals are raised} such that the operation performs reliably in our simulations. However, other timing values can also perform the same function, e.g., \upla{} performs the same function by raising the \emph{wl} signal at 4 ns, and the \emph{EQ} signal at 8 ns.}

Our SPICE simulations show that \upla{} consumes the same power independently of the initial value of the cell (as the final value is always $V_{dd}$/2). 
\om{Section~\ref{sec:emulating_upla} evaluates the quality of \upla{} \omm{values (i.e., the quality of \mechanism{}-sig PUF)} in real off-the-shelf DRAM  chips.}

\subsubsection{\dtran{}}
\label{sec:dtran}
\dtran{} generates deterministic values. The key idea is to drive the cell to a deterministic value by activating the two signals that drive the \onur{SA} (sense\_n and sense\_p) with a delay between them. Depending on which of the two signals triggers first, the generated value is 0 or 1.

Figure~\ref{fig:fleco_commands}\om{b} shows the SPICE simulation of \dtran{} generating a zero value. \dtran{} triggers the \emph{sense\_n} signal (7 ns), deviating the bitline voltage towards zero. After some delay, \dtran{} triggers the \emph{sense\_p} signal (14 ns). When both signals triggered at the same time, the SA behaves as a regular activation (see Section~\ref{sec:background}), sensing and amplifying the bitline voltage deviation introduced previously. The value generated with this sequence of signals is deterministic, and always zero. Similarly, for generating a one value, \dtran{} triggers the \emph{sense\_p} (7 ns) signal first, and the \emph{sense\_n} later (14 ns) (not shown in \om{Figure~\ref{fig:fleco_commands}b}).

\subsubsection{Other \mechanism{} Variants}
\label{subsubsc:other_variants}


The combination of different timings for triggering/disabling the four control signals enables the implementation of \ommmm{$300^4$} different variants of the \mechanism{} command. For each signal, there is a total of $n=300$ valid combinations of timings.\footnote{To calculate the number of variants $n$, we consider that the time step is $s=1ns$, and the time window of \mechanism{} is $w=25ns$. We sum the number of variants for all signals that start at different timings, i.e., there are 24 variants that trigger at 0ns (the end time can be at 1ns, 2ns... 24ns), 23 variants that trigger at 1ns (the end time can be at 2ns, 3ns... 24ns), etc. Therefore, the total number of different variants is  $n = \sum_{i=1}^{w-1} i$, which is 300 when $w=25$.} For a total of $r=4$ signals, there are \om{$t=300^4$} valid \mechanism{} variants. The large number of possible combinations provides an enormous design space for optimizing existing DRAM commands and proposing new research ideas. For example, we can use the results of error-characterization studies to optimize a particular DRAM command. To do so, we can re-implement the command using a \mechanism{} variant with timing signals optimized for reliability, performance or energy for a particular DRAM chip.

The number of \mechanism{} variants that result in fundamentally different functionalities is more limited, as the functionality of a particular \mechanism{} command is determined by the relative order in which the internal circuits are triggered and deactivated. During our exploration of the substrate, we have identified other \mechanism{} variants with different properties. For example, by triggering the \emph{sense\_p} and \emph{sense\_n} signals while the bitline is precharged (i.e., without triggering the \emph{wl} signal), we can create \lois{signatures that depend on process variation} \lois{\emph{without}} destroying the content of the memory cells, unlike \upla{}. We \onur{do} not include the evaluation of this variant because it \lois{requires changes to} commodity DRAM chips, \lois{which makes it very challenging to evaluate in real DRAM chips}.


\subsection{Implementation Details}
\label{sec:implementaion_details}

Implementing the \mechanism{} substrate in commodity DRAM chips requires \emph{minimal} changes to (1)~the circuit that generates internal timing signals to control internal components in DRAM, and (2)~the DRAM interface.
\subsubsection{\mechanism{} Circuit Design} 
\label{sec:circuit_design}
\begin{figure}[h] \centering
    \includegraphics[width=0.75\linewidth]{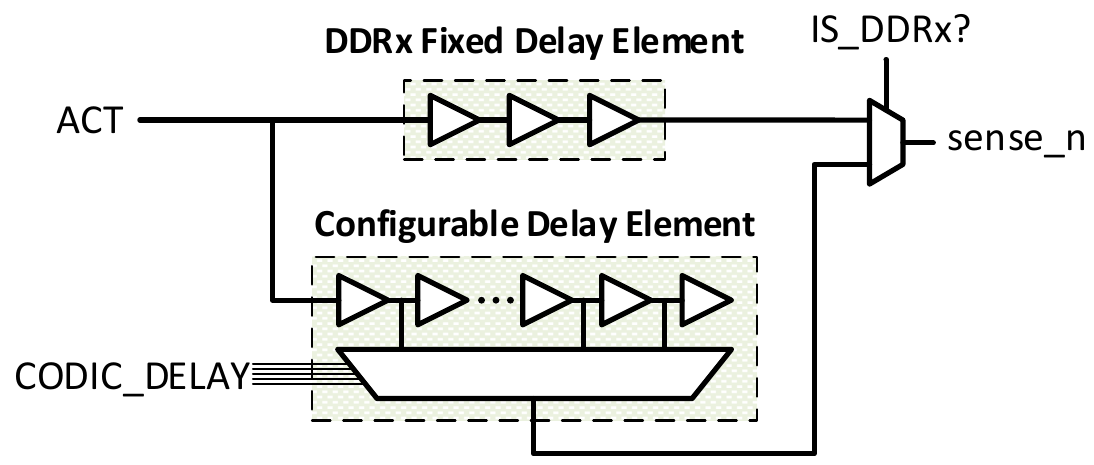}%
    \caption{\onur{Configurable} \mechanism{} delay path vs. \onur{fixed} DDRx \om{ACT command} delay path \om{for generating the sense\_n \omm{internal} signal}.}
    \label{fig:codic_delay_path}
\end{figure}

To implement the fine-grained configurable control of the timing interval between DRAM internal circuit control signals, we design a configurable delay element using a chain of buffers and a multiplexer, as Figure~\ref{fig:codic_delay_path} shows.

\setlength\parfillskip{0pt plus .4\textwidth}
\setlength\emergencystretch{.1\textwidth}
We model the delay element using the 22nm PTM transistor model~\cite{PTM} in SPICE. \lois{Each} buffer stage has a propagation delay of approximately 1ns at the output of the 25-to-1 multiplexer. We estimate the area cost of the circuit by summing up the area of the transistors used for all the buffers and the multiplexer. Assuming a $6F^2$ DRAM cell size~\cite{Roeth20156F2, Schloesser20086F2}, the area overhead of a configurable delay circuit of \mechanism{} is about 0.28\% \lois{per mat (a typical mat is 512 rows by 512 columns)}. The total hardware area overhead for controlling all four DRAM internal control signals (i.e., \emph{wl}, \ommmm{\emph{EQ}}, \emph{sense\_n}, and \emph{sense\_p}) is $4\times 0.28\% = 1.12\%$ \om{per mat.}\footnote{\omm{If needed, we can reduce the area overhead by coarsening the granularity of time control in a CODIC command.}}
The energy overhead of the \mechanism{} circuit design is less than 500fJ, which is negligible compared to the energy consumption of an activation command ($\approx$ 17nJ).
The 2-to-1 multiplexer introduces a negligible delay (0.028ns) \lois{to the DDRx activate command}, but we compensate for this by sizing the buffers such that the total delay of \lois{an activate command} is still the same as in \ommm{the} conventional DDRx \ommm{interface}.

The additional logic required to implement \mechanism{} does not affect the operation of common \ommm{DRAM} commands for two reasons: (1)~\mechanism{} does \om{\emph{not}} require any changes to the DRAM array, and (2)~\ommm{DRAM} commands still use the same logic to generate the internal circuit timings as they do in commodity DRAM chips. For example, as Figure~\ref{fig:codic_delay_path} shows, 
\onur{the IS\_DDR\om{x} signal selects between the fixed delay element for an activation command and the \mechanism{} configurable delay element.} 
A DDRx \lois{activate} command still uses the conventional fixed delay element to generate a \texttt{sense\_n} signal with proper delay after the \texttt{wl} signal.

\subsubsection{\mechanism{} Interface} 
\label{sec:interface}
We propose adding a new DRAM command to the DDRx interface to leverage the \mechanism{} substrate using a similar approach to prior academic \om{work~\cite{seshadri2017ambit,seshadri2013,Chang2016LISA,wang2020figaro,hassan2019crow,lee2019twice, bhati15Flexible, Kim:2012,Hassan2016,wang2018reducing}} and patents~\cite{greenfield2016throttling,bains2016distributed}. The new \mechanism{} command has the same format as a regular activation. We can integrate the new command in the JEDEC standard specification~\cite{jedec2012jedec} without extra cost, as there is reserved space for including new commands. By adding a single \mechanism{} command to the DDRx interface, we can implement many different variants of the CODIC command by programming 4 dedicated 10-bit mode registers (MR\om{s}) in DRAM, which are used to store the timings of the 4 internal timing signals that \mechanism{} can modify.  \mechanism{} uses the existing mode register set (MRS) command defined in the DDRx specification to change the contents of its MR.\footnote{\lois{JEDEC standard uses mode registers to dynamically change certain DRAM parameters such as burst length, burst type, CAS latency, or DLL reset~\cite{jedec2012jedec}}.} To support more than one \mechanism{} variant at the same time, or to allow independent DRAM internal circuit timing configurations for different regions of DRAM, the manufacturer can incorporate more MR registers dedicated to \mechanism{}.

\subsection{\mechanism{} Latency and Energy Evaluation}
\label{sec:eval_lat-ener-area}

\lois{In this section, we study the latency and energy of different variants of the \mechanism{} command.}

\vspace{3pt}\noindent\textbf{Methodology.}
We calculate energy consumption by using the activate and precharge energy consumption described in the power model of the DRAMPower simulator~\cite{chandrasekar2012drampower}.

\vspace{3pt}\noindent\textbf{Latency and Energy Results.}
Table~\ref{table:energy_latency} shows the latency and energy of five \mechanism{} variants that can be implemented with \mechanism{}. We show two \mechanism{} variants that mimic the behavior of regular \onur{activat\om{ion}} \onur{(\mechanism{}-activat\om{e})} and precharge \onur{(\mechanism{}-precharge)} commands, our proposed \mechanism{} variants (\upla{}, and \dtran{}), and the optimized \upla{} variant (\upla{}-opt) explained in Section~\ref{sec:u-pla}.

\begin{table}[t]
    \caption{Latency and energy of five \mechanism{} command variants.}
    \vspace{-0.1cm}
    \centering
    \begin{threeparttable}
    \footnotesize{}
    \begin{tabular}{ r  c  c }
      \toprule
         {\emph Primitive } & {\emph Latency (ns)} & {\emph Energy (nJ)} \\
        \midrule
        \onur{\mechanism{}-activate} & 35 & 17.3 \\
        \onur{\mechanism{}-precharge} & 13 & 17.2 \\
        \textbf{\upla{}} & 35 & 17.2 \\
        \textbf{\upla{}-opt} & 13 & 17.2 \\
        \textbf{\dtran{}} & 35 & 17.2\\
        \bottomrule
    \end{tabular}
    \end{threeparttable}
    \label{table:energy_latency}
\end{table}

We make two major observations about the latency results. First, \upla{}-opt is significantly faster than \upla{}, and \dtran{}. However, the \upla{}-opt and \upla{} latency does not include the additional activation command needed to \om{amplify the DRAM cells to zero or one depending on process variation} (see Section~\ref{sec:u-pla}). Second, the latency of \upla{} and \dtran{} is \onur{similar to} a standard \om{DDRx ACT} command. 

Our main observation about the energy results is that they are very similar across all \mechanism{} variants we evaluate for two main reasons. First, all the evaluated variants of \mechanism{} need to route the address within DRAM, which is one of the main sources of energy consumption \om{of all commands} (around 40\%~\cite{chandrasekar2012drampower}). Second, the energy consumption of the sense amplifier (used in \upla{}) and the precharge logic (used in \upla{}) are similar \om{across all commands} (around \om{40\%~\cite{chandrasekar2012drampower}}). 

\subsection{Limitations and Challenges}
\label{sec:limitations_challenges}
\mechanism{} is a substrate that can potentially enable many applications, which \lois{might} introduce \lois{application-specific} implementation challenges. For example, one of the main challenges to consider when using \mechanism{} PUFs \onur{(Section~\ref{sec:physicallyUnclonable})} is that \mechanism{} works at the granularity of a DRAM row, which could contain data from multiple virtual memory pages (e.g., critical data or data from another process that should not be destroyed). Previous works\onur{~\cite{seshadri2017ambit, seshadri2013, Seshadri:2015, Chang2016LISA, hajinazar2021simdram}} \ommm{propose} ideas that have similar \omm{challenges}, and they propose solutions that \ommmm{also} could be \om{used with} \mechanism{}.

One \onur{general} concern is that exposing internal signals to users may pose potential security concerns.
To preserve \mechanism{} functionality without exposing the internal signals, we can instead provide a more controlled and isolated interface.
This interface would provide users with commands to directly invoke the new \mechanism{} applications,
instead of allowing the user to access the internals of DRAM directly. 
For example, the memory controller could provide a direct command for \mechanism{} PUF \onur{(Section~\ref{sec:physicallyUnclonable})} to the user (as either a software API call or as a new assembly instruction). Internally, the controller would keep track of a system-defined memory address range that is safe to use to generate a PUF response, and the controller would internally use \mechanism{} to control the DRAM timings and generate the PUF response.
While such an approach prevents user-generated \mechanism{} applications, we can still benefit from new \mechanism{}-based applications that are implemented and exposed by \onur{memory controller and} CPU manufacturers.

%% file: 05_applications.tex
\section{Applications Enabled by \mechanism{}}
\label{sec:otheraplications}
\mechanism{} can be used in many applications. As a proof-of-concept, we design two new applications in the security domain that outperform the best state-of-the-art mechanisms (Section~\ref{sec:physicallyUnclonable} and Section~\ref{sec:coldbootattacks}). We also discuss other possible applications that can be implemented with \mechanism{} (Section~\ref{sec:other_apps}).

\subsection{Physical Unclonable Functions (PUFs)}
\label{sec:physicallyUnclonable}

A \om{PUF~\cite{Gassend:2002,Daihyun2005,Suh2007,kim2018}} is a hardware primitive that maps \omm{a unique} \lois{input (i.e., challenge)} to a \lois{unique} response. \lois{A response is} typically derived from the unique physical characteristics  (resulting from process variation) of an integrated circuit, such that no two circuits can provide the same response to the same challenge. A PUF can be used as a building block for implementing low-cost authentication protocols~\cite{Majzoobi2012, Hammouri2008, Rostami2014, Che2015}, key generation applications~\cite{Roel2012,Yu2012,Paral2011}, \onur{intellectual property protection~\cite{guajardo2008brand,zheng2014digital,guajardo2007fpga}, \om{hardware obfuscation~\cite{wendt2014hardware,khaleghi2018hardware}}, prevention of \om{reverse engineering~\cite{wei2014reverse,Rostami2014}}, intellectual property watermarking~\cite{bordel2019digital}, software metering~\cite{dabiri2009hardware}, or prevention of hardware Trojan embedding~\cite{wendt2013bidirectional}}.

In the specific case of DRAM-based PUFs, one or more parameters \onur{(e.g., address and size of a memory segment)} define a challenge, and the data read from DRAM is the response to that challenge. Together, they define a Challenge-Response pair (CR pair). In this work, we use the address \lois{and size} of \lois{a memory} segment as the only \lois{parameters that define} a challenge.

\vspace{3pt}\noindent\textbf{Limitations of State-of-the-Art DRAM-based PUFs.}
\label{sec:stateoftheart_PUFs}
Prior DRAM-based PUF proposals exploit variations in DRAM start-up values~\cite{Tehranipoor:2015}, DRAM write access latencies~\cite{Hashemian:2015}, DRAM cell retention failures~\cite{Sutar2016, Xiong2016, Keller2014} and reduced DDRx timing parameters~\cite{kim2018,talukder2019prelatpuf}. There are \lois{at least four} main limitations with most of these approaches. 
\lois{First}, most DRAM PUFs~\cite{Tehranipoor:2015,Hashemian:2015,Sutar2016, Xiong2016, Keller2014} have long evaluation times, which may incur significant system interference when the PUF is evaluated at runtime. 
\lois{Second}, most DRAM PUFs~\cite{Tehranipoor:2015,Hashemian:2015,Sutar2016, Xiong2016, Keller2014,kim2018} require heavy filtering mechanisms to deal with the \onur{inherently} noisy nature of the DRAM responses. \ommmm{Filtering mechanisms increase the reliability of PUF responses at the cost of increasing} evaluation latency. 
\lois{Third}, the responses to the same challenge in most DRAM PUFs~\cite{Tehranipoor:2015,Hashemian:2015,Sutar2016, Xiong2016, Keller2014,kim2018} exhibit high variation with temperature changes, which is an important issue in systems with a non-controlled environment (e.g., IoT devices in the wild). 
\lois{Fourth}, some DRAM PUFs are data dependent, which may cause \onur{the same challenge to have different} responses \onur{depending} on the content of the memory~\cite{Sutar2016, Xiong2016, Keller2014,kim2018,Tehranipoor:2015,Hashemian:2015}. 

\vspace{3pt}\noindent\textbf{\mechanism{}-based PUFs.} 
\upla{} can generate \lois{signature values that depend on process variation} \onur{and thus} can be used as PUF \lois{responses} (Section~\ref{sec:substrate}). \UPLAPUF{} provides \lois{four} main advantages compared to the state-of-the-art DRAM PUFs. 
\loiss{First, \upla{} has a fast evaluation time due to its ability to control internal DRAM timing signals. The absence of a filtering mechanism combined with the short latency of \upla{}, enables the \UPLAPUF{} to have a shorter evaluation time than the best state-of-the-art DRAM PUF~\cite{talukder2019prelatpuf}. 
Second, in contrast to other DRAM PUFs, \upla{} does not require any filtering mechanisms because it provides highly stable output values. This stable output allows computing a valid PUF response with a simple low-overhead filter, or without any filter at all (see Section~\ref{sec:eval_randomnes}), depending on the requirements of the application.} 
Third, \upla{} has state-of-the-art resilience to temperature changes, i.e., changing the temperature does not influence much the repeatability of the PUF responses. 
\lois{Fourth, \upla{} \ommmm{responses} do not depend on the content of DRAM, \ommmm{as all DRAM cells are always precharged to $V_{dd}/2$ for generating a PUF response, independently of their original value.}}

\subsection{Preventing Cold Boot Attacks}
\label{sec:coldbootattacks}

A cold boot attack is a physical attack on DRAM that involves hot-swapping a DRAM chip and reading out the contents of the DRAM chip on another \om{system~\cite{Yitbarek2017,Halderman2009,Simmons2011,Gruhn2013,BAUER2016S65,muller2010aesse,villanueva2019cold,mcgregor2008braving,lindenlauf2015cold,lee2011correcting,Hilgers2014}}. The attacker first disables power to the computer containing the victim DRAM and then transfers the DRAM to another system that can read its content. Alternatively, the attacker can also recover the memory contents by booting a small special purpose program from a cold reset \onur{(i.e., reset after \om{power cycling)~\cite{Halderman2009}}}. 
Cold boot attacks are possible because the data stored in DRAM is not immediately lost when the chip is powered-off. This is due to the capacitive nature of DRAM cells that can hold their data up to several \om{seconds~\cite{BAUER2016S65, Patel:2017, Liu:2012, Liu2013, khan2014efficacy}} or minutes~\cite{Halderman2009}. This \onur{data retention} effect can be even more significant if the DRAM module is swapped at low temperatures. For this reason, protection mechanisms such as disabling refresh on DRAM for a certain amount of time are not effective at protecting against cold boot attacks.

There are three classes of mechanisms for preventing cold boot attacks.
First, mechanisms that rely on encrypting memory either explicitly~\cite{Suh:2003, arnold2004ibm, Yang:2005, Duc2006, Rogers2007, Henson:2014, amd_sme, Yitbarek2017}, or implicitly through some CPU extensions (e.g. Intel SGX~\cite{costan2016intel}, AMD SEV~\cite{amdsev}). These mechanisms are effective and secure, but \onur{are} complex and expensive (in terms of energy and performance overhead) to be implemented in many low-cost devices.
Second, modern systems scramble the data in the memory \lois{controller}, which helps to obscure the DRAM contents. This mechanism is simple and is also \onur{useful} for other purposes (e.g., improving signal integrity on the DRAM bus), but it has been shown to be insecure against cold boot attacks~\cite{Yitbarek2017}.
Third, the mechanism proposed by the Trusted Computing Group (TCG)~\cite{computing2009tcg} \lois{resets} the DRAM content upon power-off. 
This mechanism is implemented on the host platform firmware and depends on the OS, which makes it vulnerable to attacks~\cite{fsecureattack}. 

We identify at least two scenarios where a fast and secure boot is critically important.  First, devices that run intermittently (intermittent computing)~\cite{lucia2017intermittent} extract energy from their environment (e.g., solar energy), and they \ommmm{power-on} and execute a burst of operations only once they have enough energy. These devices usually rely on volatile memories such as DRAM~\cite{lucia2017intermittent}. Such devices have very limited energy available to them, and stalling for even a few seconds can reduce the amount of compute possible (and can potentially trigger a greater number of rollbacks).
Security is also critically important,  because these devices are usually in the wild, and are vulnerable to physical attacks such as cold boot attacks. Second, devices that require instant-on capabilities~\cite{chang2006instant,lee2010instant,stenger2008wildlife}, such as unattended in-the-wild digital cameras that need to \ommmm{power-on} instantly when detecting a signal for automatically taking a picture~\cite{stenger2008wildlife}.

\subsubsection{Threat Model}
\label{sec:threatModel}

We assume a threat model in which the attacker has physical access to a live uncompromised machine/device for an unlimited amount of time to steal information stored in the device's DRAM. 
In order to successfully read data from DRAM, the attacker cannot keep the DRAM \ommmm{powered-off} for long, as this would lead to data loss.
We therefore assume that, as part of the attack, the DRAM chip is \ommmm{powered-off} for an arbitrarily short amount of time. This power loss occurs either (1)~when transplanting the DRAM module to an attacker-controlled machine, or (2)~during attacks that reboot the victim machine to load a malicious OS. 
Note that our threat model does not account for:
(1)~systems with a warm reboot that never cuts power to DRAM, or
(2)~hypothetical attacks using a scanning electron microscope (SEM) to read DRAM content.%
\footnotemark[\getrefnumber{foot:threat_model}]
\addtocounter{footnote}{1}\footnotetext{\label{foot:threat_model}Note that to our knowledge, this attack has not been demonstrated to date on commodity DRAM chips, but only under a controlled environment using custom DRAM samples~\cite{sem}.}

\subsubsection{Self-Destruction}
\label{sec:self-destruction}

We make the observation that it is possible to protect from cold boot attacks by deleting all memory contents during the DRAM power-on. Based on this observation, we propose \emph{self-destruction}, a low-cost in-DRAM mechanism based on \mechanism{} that destroys all DRAM contents without the intervention of the memory controller during DRAM power-on.

The key idea of self-destruction is to destroy the entire DRAM memory using \upla{} or \dtran{}. Self-destruction is performed autonomously without the intervention of the memory controller, which could be controlled by the attacker.
During self-destruction, the DRAM chip does not accept any memory commands to ensure the atomicity of the process.
On a cold boot reset, our mechanism steps through a sequence of \mechanism{} commands in order to destroy the contents of the entire chip sequentially (i.e., row-by-row).

\vspace{3pt}\noindent\textbf{Hardware Implementation.}
We propose two ways to implement self-destruction within DRAM. First, we propose to add a dedicated circuit that (1)~issues \mechanism{} commands back-to-back to all DRAM rows, (2)~parallelizes commands across banks, and (3)~enforces JEDEC timing constraints (e.g., t\_FAW~\cite{jedec2012jedec}). We evaluate this implementation in Section~\ref{sec:eval_coldBootAttacks}.
Second, we propose to reuse the existing circuits for \om{self-refresh~\cite{lpddr4x}}  to issue \mechanism{} commands in our self-destruction mechanism. This implementation is an optimization for reducing the hardware cost, but the destruction time is the same as the time that the self-refresh mechanism takes to refresh the entire memory.

Implementing self-destruction by reusing self-refresh circuits requires two small modifications to the self-refresh circuitry that enhance its default functionality (i.e., self-refresh) to additionally enable it to destroy the contents of DRAM at power-on. First, we add configurable delay circuits (see Figure~4) that generate internal DRAM signals (i.e., wl, sense\_p, sense\_n, and EQ in Figures~2 and~3) with configurable timings. These newly-generated signals are used to implement the CODIC operations within the self-refresh circuitry. Second, for each connection between an internal DRAM signal (i.e., wl, sense\_p, sense\_n, EQ) and the DRAM array, we add a MUX that selects whether to use (a) the unmodified signals generated by the self-refresh circuitry for the baseline refresh operation, and (b) our newly-generated signals for enabling data destruction using CODIC. The select line of each MUX is set at \ommmm{power-on} (e.g., to 1) to choose the CODIC signals and destroy the DRAM contents, and the select line's value changes (e.g., to 0) immediately after the entire contents of DRAM are destroyed to enable self-refresh to be used for the rest of the time.

Any implementation of self-destruction also needs logic to trigger the self-destruction mechanism at DRAM power-on. 

\vspace{3pt}\noindent\textbf{Security Analysis.}
Our mechanism is automatically triggered in DRAM when power is detected, without requiring external action. Therefore, the security of self-destruction depends on the reliability of the DRAM module's power-on detection circuit. There are two ways for an attacker to potentially bypass this circuit. We describe both and explain why, in practice, they do not pose a security threat. 

First, an attacker could operate DRAM at low voltage on the compromised system using, for instance, Dynamic Voltage and Frequency Scaling (DVFS)\onur{\cite{deng2011memscale,Chang:2017,david2011memory,haj2020sysscale}}, with the goal of \emph{not} triggering the power-on detection circuit. The power-on circuit triggers when it detects a voltage ramp up from $0V$, but \onur{the voltage} does not need to reach $V_{dd}$ (\onur{the power-on circuit} triggers as long as a voltage ramp up starting from $0V$ is detected). Therefore, operating the DRAM at very low voltage would not help the attacker.\footnote{The attacker might try to operate the device at a voltage close to $0V$ such that the power-on circuit cannot detect the ramp up, however, at such low voltage the DRAM \om{chip} would not be operational.}

Second, an attacker could \onur{overheat} the DRAM power-on detection mechanism \onur{until it stops functioning}. In practice, however, the FSM that initializes the chip is in the same internal controller that regulates other functions (i.e., timing signals for activate, precharge, and other commands). Consequently, \onur{overheating} that component would most likely make the whole DRAM unusable.

\subsection{Other Applications}
\label{sec:other_apps}

The \mechanism{} substrate enables the implementation of a wide variety of applications. We identify and classify several potential applications into three broad categories, though we note that there may be other applications or categories that can also make use of \mechanism{}.

\subsubsection{Security Applications}
\label{sec:app_sec}

In \onur{Sections~\ref{sec:physicallyUnclonable} and~\ref{sec:coldbootattacks}}, we propose and evaluate in detail new mitigations mechanisms for two security applications that improve over state-of-the-art. First, \mechanism{} can implement two new Physical Unclonable Functions (PUFs) that improve upon state-of-the-art DRAM PUFs (Section~\ref{sec:physicallyUnclonable}). Second, \mechanism{} enables the implementation of a new mechanism \onur{called self-destruction} to \lois{protect} against cold boot attacks (Section~\ref{sec:coldbootattacks}). 

\mechanism{} can also be used to accelerate secure deallocation mechanisms~\cite{chow2005shredding, Anikeev:2013, Sha2018, harrison2007protecting, Garfinkel:2004}. Secure deallocation is  a  technique  that  sets the data to zero at the moment of deallocation, or as soon as the data is not needed anymore, which reduces the time that critical data is exposed to attacks. We evaluate our mechanism and demonstrate that it performs up to 21\% better and consumes 34\% less energy than secure deallocation techniques implemented in software.\todo{double check the numbers}

\mechanism{} \lois{also} enables accurate control of the failure mechanisms used in \lois{existing} DRAM-based \ommmmm{True Random Number Generators (TRNGs)~\cite{kim2019d, olgun2021quactrng, talukder2019exploiting}} and DRAM PUFs~\cite{kim2018,talukder2019prelatpuf}. These mechanisms are triggered by violating the DRAM standards, but the exact internal failure mechanism is not \om{perfectly} known (e.g., it might vary depending on the DRAM architecture, timings, manufacturer). A substrate such as \mechanism{} would help to understand and improve the quality and \om{performance} of these \ommmmm{existing} mechanisms, \ommmmm{and enable new TRNGs that exploit new failure mechanisms for generating random numbers.}

\ommmmm{\mechanism{} could also potentially be used to reduce the effectiveness of RowHammer attacks~\cite{Kim2014,kim2020revisiting,mutlu2019rowhammer,frigo2020trrespass,seaborn2015exploiting}. Because \mechanism{} allows finer-granularity control over DRAM timings, a system designer can potentially adjust the timings of key internal DRAM signals that govern important parameters that affect the susceptibility of a chip to RowHammer, e.g., wordline active time~\cite{jiangquantifying}.}

\subsubsection{Custom DRAM Latency/Energy Optimizations}
\label{sec:app_optimization}

Existing DRAM chips are shipped with fixed and conservative internal circuit timings that guarantee \lois{reliable} behavior for a wide range of environmental conditions and process variation. Using \mechanism{}, the internal circuit timings can be optimized for a particular DRAM device, based on the device's current environmental conditions. 

\noindent\textbf{Adapting to Environmental Conditions.} \mechanism{} enables dynamic and practical adaptation of DRAM internal circuit timings to different environmental conditions (e.g., temperature, voltage, aging), with the goal of optimizing DRAM performance and energy consumption.

\noindent\textbf{Mitigating Process Variation Effects.}
\mechanism{} enables practical adaptation of DRAM internal circuit timings to process variation of a particular DRAM chip. For example, the retention time of DRAM cells depends on process \omm{\mbox{variation}\onur{~\cite{Liu2013, BAUER2016S65, Patel:2017, Liu:2012, khan2014efficacy, Halderman2009}}}. \mechanism{} can easily provide an alternative activate command for rows with high retention times, which reduces the time between triggering the wordline and triggering the sense amplifier. Using this technique, we enable rows with high retention times to be activated faster \om{during refresh}.

\noindent\textbf{\onur{Accurate DRAM} Characterization.} 
\mechanism{} enables increased precision and opportunities in DRAM characterization. For example, by modifying the time interval between triggering the \emph{wl} and the \emph{sense\_p and sense\_n} signals in an activation command, we can test how fast the cell capacitor shares its charge with the bitline, and infer the relative capacitance of each cell. Such observations would enable us to build more features with a deeper and \om{more accurate} understanding of their underlying mechanisms. \onur{They can also enable us to build more accurate models of DRAM behavior.}

\noindent\textbf{\lois{Memory Controller Timing Parameters.}} \mechanism{} can be used to reliably enforce the usage of reduced \lois{memory controller} timing parameters (e.g., tRCD), 
\lois{as \onur{it enables} us to know and control the precise state of one command when the next command is issued with reduced timing parameters. }
This \om{can better} enable prior \omm{works~\cite{Lee2015,Hassan2016,Chang:2016,zhang2016restore,Lee:2017,wang2018reducing, kim2018solar,koppula2019eden}} to \onur{optimize DDRx timing parameters while ensuring \om{robust} operation, as the memory controller can make informed decisions about reducing DDRx timing parameters taking into account the DRAM internal state, which is known and can be controlled with \mechanism{}.}

\subsubsection{Processing-in-Memory (PIM)}
\label{sec:app_existing}
PIM proposals such as RowClone~\cite{seshadri2013} \onur{and Ambit \om{(bitwise AND, OR)}~\cite{seshadri2017ambit,seshadri2016buddy,seshadri2019dram}} have been demonstrated in real \om{off-the-shelf} DRAM chips via manipulating DDRx timing parameters~\cite{gao2019computedram}. 
\onur{However, the results shown in \cite{gao2019computedram} demonstrate that \om{many intended computations} cannot be performed reliably in \om{a} vast majority of \om{chips and cells}} 
because it is not possible to control how DRAM commands are executed internally within DRAM chips. An interface like \mechanism{} can \om{more robustly} enable RowClone\om{, Ambit,} and other PIM \omm{mechanisms~\cite{hajinazar2021simdram}} on real DRAM chips.

%% file: 06_evaluation.tex
\section{Evaluation}

We evaluate the quality, latency, and randomness of \om{the} \mechanism{} PUF  \lois{(Section~\ref{sec:emulating_upla})}, and the latency and energy of our cold boot attack prevention mechanism \lois{(Section~\ref{sec:eval_coldBootAttacks})}.

\subsection{\mechanism{} PUF}
\label{sec:emulating_upla}
\label{sec:eval_randomnes}


We evaluate the quality of the \ommmm{\mechanism{} PUF (i.e., \mechanism{}-sig PUF)} responses with a new methodology that allows us to recreate the \lois{PUF responses} of the  \UPLAPUF{} in real \om{off-the-shelf} DRAM chips.
We perform an exhaustive evaluation using 136 \emph{real} \ommmm{DDR3} DRAM chips from 15 modules.

\vspace{3pt}\noindent\textbf{Methodology.}
A \upla{} primitive (1)~sets a cell to $V_{dd}/2$ with the precharge logic, and (2)~activates the SA to generate \onur{a} \lois{signature} value from that cell (as explained in Section~\ref{sec:u-pla}).
We emulate this behavior in \emph{real} DRAM chips in two steps. First, based on the observation that a DRAM cell leaks towards $V_{dd}/2$, we disable \onur{DRAM} refresh for 48 hours with the goal of setting the cell to $V_{dd}/2$. Second, we activate this cell to obtain the PUF response. This methodology allows us to reproduce the responses that would produce a real \UPLAPUF{} implementation. Recall that discharging the cells \om{with \UPLAPUF{}} would take a few nanoseconds (\emph{not} 48h) in a real implementation (Section~\ref{sec:eval_lat-ener-area}). We perform our experiments with a customized memory controller built with SoftMC~\cite{hassan2017} and a Xilinx ML605 FPGA on 136 different DDR3 DRAM chips from three major vendors.
Table~\ref{table:dram_chips_short} shows the main characteristics of the 136 DRAM chips that we evaluate. \loiss{Appendix~\ref{sec:DRAMchips} has more details.}

\begin{table}[h]
    \caption{Characteristics of the 136 evaluated \om{DDR3} DRAM chips.}
        \vspace{-0.1cm}
    \centering
    \footnotesize{}
    \setlength\tabcolsep{3pt} 
    \begin{tabular}{ c  c  c  c  l   }
        \toprule
        {\bf Vendor} & \loiss{{\bf Chips}} & {\bf \ommm{Capacity}/chip} & \loiss{{\bf Freq. (MT/s)}} &  \multicolumn{1}{c}{\bf Voltage}  \\
        \midrule
        A & 32 &  \om{4Gb} & 1600 & 1.35V (DDR3L) \\ 
        A & 32 &  \om{4Gb} & 1600 & 1.50V (DDR3) \\ 
        B & 32 &  \om{2Gb} & 1333 & 1.50V (DDR3) \\ 
        B & 8 &  \om{4Gb} & 1600 & 1.35V (DDR3L) \\ 
        C & 32 & \om{4Gb} & 1600 & 1.35V (DDR3L) \\ 
        \bottomrule
    \end{tabular}
    \label{table:dram_chips_short}
\end{table}

\onur{Evaluating} the \upla{} PUF functional behavior with our methodology is challenging, because DRAM cells can retain their content for a long time\onur{~\cite{Liu2013,Liu:2012,venkatesan2006retention,Patel:2017,khan2014efficacy}}, i.e., not refreshing the DRAM does not guarantee that a cell will end up with the precharge voltage ($V_{dd}/2$), even after a long period.
 
To deal with this issue, we tailor a custom test to determine if a cell is set to the precharge voltage. As discussed in Section~\ref{sec:u-pla}, when a cell is set to the precharge voltage, the value that \upla{} generates should always \onur{be} the same regardless of the initial value of the cell. Based on this observation, our test analyzes the final value of a DRAM cell after 48 hours without refresh, in two different scenarios:
(1)~all initial values are zero, and (2)~all initial values are one. 
The test has two possible outcomes.
First, the \emph{test passes} if the final value is the same regardless of the initial value. Thus, we can conclude that the cell is set to the precharge voltage. 
In this case, the final value is the same as the value that a real \upla{} implementation would generate.
Second, the \emph{test fails} if the final value is different \om{from the initial value}. In that case, we cannot conclude that the cell is set to the precharge voltage (\onur{and thus} we cannot infer the value generated by \onur{applying} \upla{} \onur{to the cell}), so we do not consider that cell in our results.

\onur{Following this methodology, we obtain \mechanism{} values for 34\% to 99\% of all cells \om{in the 136 DDR3 \ommmm{DRAM} chips we tested}. \om{Most of these cells} are always amplified to a particular value (e.g., 0), and the remaining cells \om{(between 0.01\% and 0.22\% of all cells\footnote{We observe that these results are in line with the results obtained in our SPICE simulations \omm{(see Section 7.2.1 in~\cite{orosa2019dataplant}).}}), which are randomly spread across DRAM,} are amplified to the opposite value (e.g., 1). We use the addresses of \om{the latter type of} cells as PUF responses.}

\subsubsection{Quality \onur{Evaluation}}
\label{sec:quality_evaluation}

\onur{This section evaluates the quality of \mechanism{} PUF responses compared to \omm{two} state-of-the-art DRAM \om{PUFs~\cite{kim2018,talukder2019prelatpuf}.}}

\vspace{3pt}\noindent\textbf{\onur{Metrics.}}
To measure the \emph{uniqueness} and \emph{similarity} of a PUF, we apply Jaccard indices~\cite{jaccard1901etude} as suggested by prior works~\cite{schaller:2017, Xiong2016, aysu2017new, kim2018}. We determine the Jaccard indices by taking \lois{two sets of PUF responses from two memory segments ($u1,u2$),} and calculating the ratio of their shared values over the full set of \lois{the two PUF responses} \footnotesize $\frac{|u1 \cap u2|} {|u1\cup u2|}$\normalsize. A ratio close to 1 represents high similarity, and  a ratio close to 0 represents uniqueness.

We use the term Intra-Jaccard for representing the similarity of two \lois{responses} from the \emph{same} memory segment, and Inter-Jaccard for representing the uniqueness of two \lois{responses} from \emph{different} memory segments. An ideal PUF should have an Intra-Jaccard index close to 1 (a unique challenge has a unique response), and an Inter-Jaccard index close to 0 (different challenges have different and random responses).

\vspace{3pt}\noindent\textbf{\onur{Methodology.}}
We compute the distribution of Intra- and Inter-Jaccard indices obtained by running experiments on 
136 different DRAM chips with segments of 8KB (this size is used by prior work~\cite{kim2018}). We calculate the Intra-Jaccard indices for 10,000 random pairs of memory segments (each pair composed of two responses from the same memory segment), and the Inter-Jaccard indices for 10,000 random pairs of memory segments (each pair composed of two responses from different memory segments) from all DRAM chips.

We compare \UPLAPUF{} with the DRAM latency PUF~\cite{kim2018} and PreLatPUF~\cite{talukder2019prelatpuf}. The DRAM Latency PUF accesses DRAM with  $tRCD = 2.5ns$. For improving the repeatability of the responses, the DRAM latency PUF implements a filtering mechanism that removes cells with low failure probability from the PUF response. To this end, the mechanism reads the memory segment 100 times, and it composes a response that contains only the failures that repeat more than 90 times~\cite{kim2018}. 
\lois{PreLatPUF generates PUF responses by reducing the precharge latency ($tRP = 2.5ns$ in our evaluation)}. The authors of PreLatPUF improve their results by selecting the DRAM cells that are more suitable for PUFs via DRAM characterization. In our evaluation, we do not apply this selection mechanism, as our goal is to compare the quality of the failure \lois{mechanisms} under the same conditions. 
\om{Note} that the selection mechanism used by the authors of PreLatPUF can be also applied to the DRAM Latency PUF and \om{the} \mechanism{} PUF.

\begin{figure}[ht] 
    \centering
    \includegraphics[width=1.0\linewidth]{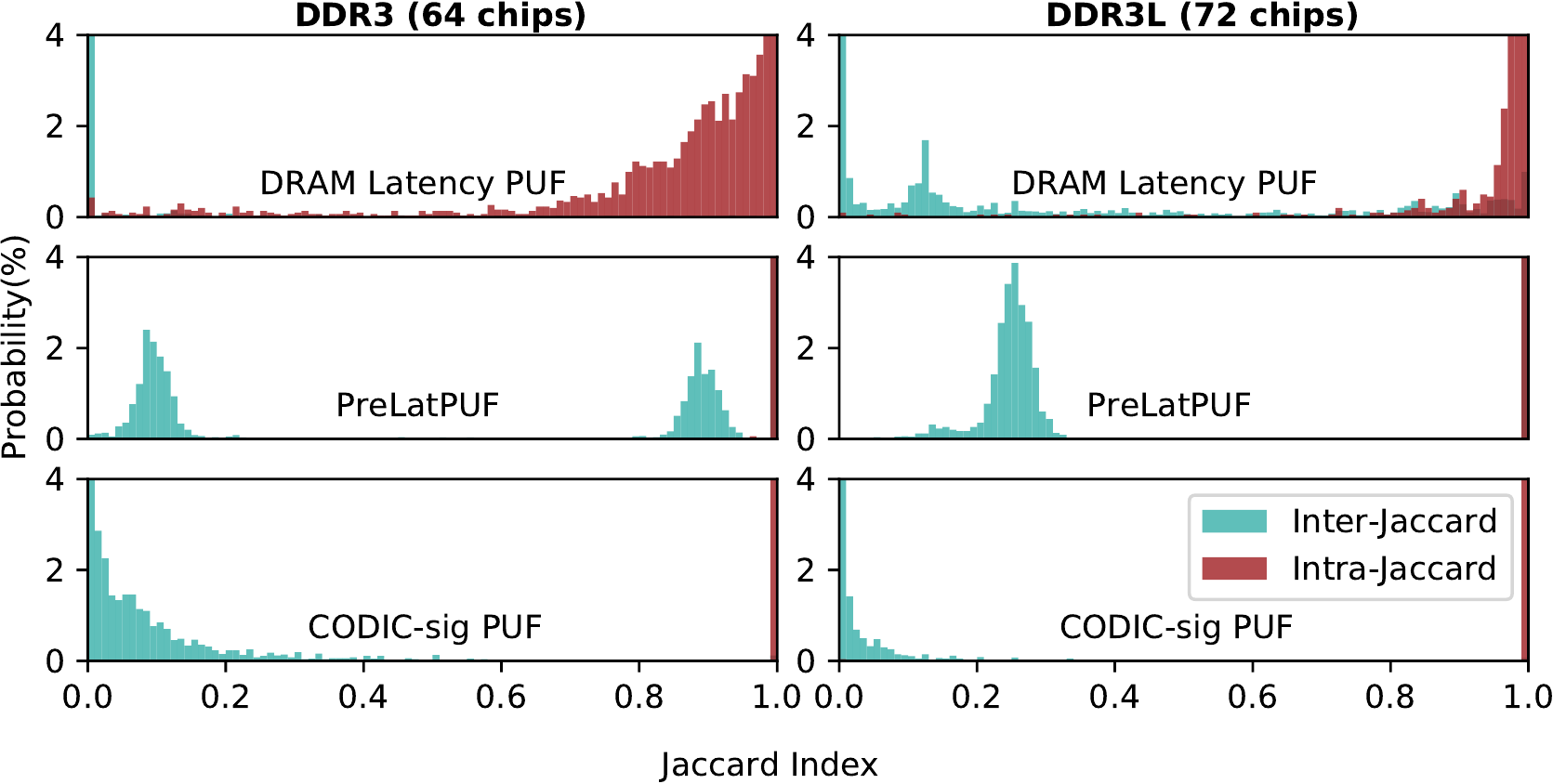}
    \vspace{-0.5cm}
    \caption{Jaccard indices obtained with the \om{the DRAM latency PUF, PreLatPUF,} and the \om{CODIC-sig} PUF, on \om{136} DDR3 and DDR3L chips.}
    \label{fig:jaccard}
\end{figure}

The \onur{PUF responses generated} by \onur{\upla{}/PreLatPUF} are less noisy than the values obtained by DRAM Latency PUF \onur{(i.e., the Intra-Jaccard indices are close to one\om{, as we will show in Figure~\ref{fig:jaccard} soon})}, so \onur{\upla{}/PreLatPUF need} a much more lightweight filtering mechanism. \onur{Although we experimentally observe} that one challenge is enough to get a robust \upla{}/PreLatPUF response in most cases \onur{(99.72\%/96.92\% of all challenges have the same response for the worst DRAM module we tested)}, \onur{we apply} a conservative filter of 5 challenges \onur{for generating \emph{always} the same response, such that \om{\upla{}/PreLatPUF are} robust even} \om{under} \onur{worst-case} conditions. While a DRAM \onur{Latency} PUF with a lightweight filtering mechanism (e.g., 1-10 reads) could be as fast as \om{the} \mechanism{} PUF, the PUF quality would decrease significantly.

\vspace{3pt}\noindent\textbf{\onur{Results.}}
Figure~\ref{fig:jaccard} shows the Intra- and Inter-Jaccard indices \om{for the DRAM Latency PUF~\cite{kim2018}, PreLatPUF~\cite{talukder2019prelatpuf},} and \UPLAPUF{} for 64 DDR3 chips operating at 1.5V and 72 DDR3L chips operating at 1.35V. 

We make four main observations. 
First, the \UPLAPUF{} shows \onur{excellent} Intra-Jaccard indices (the indices in red, which accumulate close to 1 on the x-axis), and \onur{very} good Inter-Jaccard indices (the indices in blue, which accumulate close to 0 on the x-axis). 
Second, the DRAM latency PUF has Intra-Jaccard indices distributed towards 1.0, but the distribution is still very dispersed. The Inter-Jaccard indices \om{of the DRAM Latency PUF} are very good; all values are highly concentrated around the zero value (the blue bars).
Third, PreLaTPUF shows very good Intra-Jaccard indices (the red bars near 1 on the x-axis), but Inter-Jaccard indices are very dispersed and far from the zero value. 
Fourth, the results from DDR3L chips are generally better than those from DDR3 chips, \om{especially for the \UPLAPUF{}}. We conclude that the \UPLAPUF{} is very effective \om{at providing} very similar responses to the same challenge, while maintaining uniqueness \om{across} responses \om{to different challenges}.

Based on our results, a naive challenge-response authentication mechanism implemented with \upla{} that correctly authenticates only when the response is exactly as expected (i.e., no filtering mechanism), has an average false rejection rate of 0.64\% and false acceptance rate of \ommmm{0.00\%}.

\begin{figure}[h] \centering
    \includegraphics[width=0.8\linewidth]{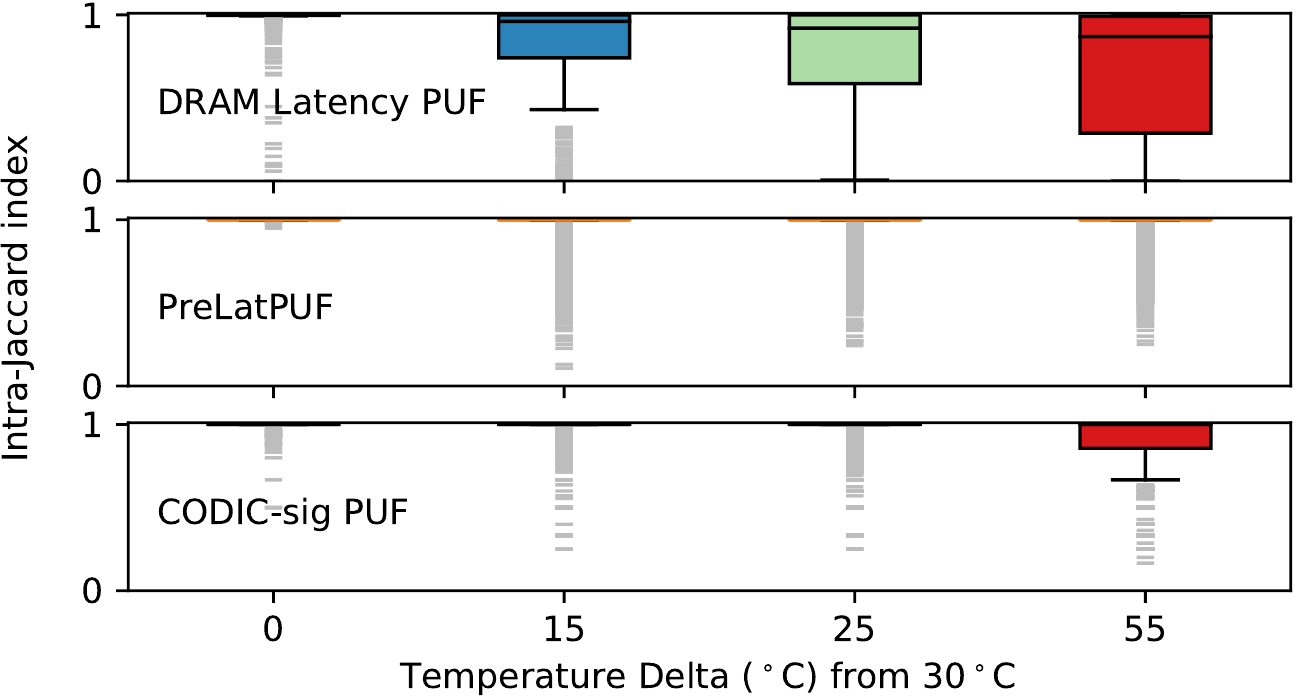}
       \vspace{-0.1cm}
    \caption{\onur{Intra-Jaccard \om{indices}} vs.\ temperature.}
    \label{fig:jaccard_temperature}
\end{figure}

\vspace{3pt}\noindent\textbf{Temperature and Aging Effects.}
\label{sec:temperature_effects}
To demonstrate how temperature affects the similarity of different responses to the same challenge. We use the experimental setup from the previous experiment \om{in Figure~\ref{fig:jaccard}}, \om{and the same 136 \ommmm{DDR3} DRAM chips}. \om{To control the temperature, we use DRAM heaters \omm{on} both sides of the DDR3 module}, and a \onur{fine-grained} temperature controller that can control the temperature with a precision of $\pm$ 0.1$^{\circ}$C. \om{We evaluate 4 different temperatures from 30$^{\circ}$C to 85$^{\circ}$C}. For this experiment, we \om{wait} for \omm{only} 4 hours (instead of 48), since cells discharge faster at high \om{temperatures~\cite{Liu2013,Liu:2012,venkatesan2006retention,Patel:2017,khan2014efficacy}}. Figure~\ref{fig:jaccard_temperature} shows the Intra-Jaccard indices between the \emph{same} segments under \emph{different} temperatures \om{across 136 chips}. 

We make three observations. First, the \upla{} PUF is very robust to temperature changes, as the responses to the same challenge are very similar even for extreme temperature changes (55$^{\circ}$C). Second, PreLatPUF has the best robustness against temperature changes, but this comes with the trade-off of having poor uniqueness properties, as the Inter-Jaccard indices show in Figure~\ref{fig:jaccard}. Third, the responses of the DRAM latency PUF are much more sensitive to temperature changes, confirming the results of the original paper~\cite{kim2018}. We conclude that the  \onur{\upla{} PUF} (1)~performs much better than the DRAM Latency PUF under changing temperature conditions, and (2)~\onur{performs} very close to PreLatPUF, which is the most robust PUF against temperature changes.

To demonstrate how aging affects the similarity of different responses to the same challenge, we use accelerated aging techniques to artificially age our DRAM chips~\cite{Sabnis,Tehranipoor,Reynolds,Saha,Sonnenfeld}. We artificially age the DRAM chips by operating them at 125$^\circ$C degrees running stress tests for 8 hours. We observe from our experiments that \upla{} PUF is very robust to aging, as most of the \onur{Intra-Jaccard} indices are 1 \onur{(not plotted)}.

\subsubsection{PUF Response Time}
\label{sec:puf_evaluation_time}

\om{Table~\ref{table:latency_PUFs} summarizes the total evaluation time of the \om{three} DRAM PUFs.} We make two observations. First, the \om{\mechanism{}-sig} PUF with/without filter \lois{has} 20x/100x lower evaluation latency than \om{the} DRAM Latency PUF. Second,  the \om{\mechanism{}-sig} PUF \onur{with/without} filter \lois{is} \onur{1.8$\times$/1.8$\times$} faster than the PreLatPUF. In our evaluation setup, although the filtering mechanism \omm{increases} \om{\mechanism{}-sig} PUF and PreLatPUF evaluation latency, it also avoids other issues related to the no-filter mechanism such as initial profiling \onur{to select robust DRAM cells}, or metadata accesses \om{required} \onur{for managing the selected DRAM cells}\onur{~\cite{talukder2019prelatpuf}}.
\onur{We conclude that the evaluation time of \onur{the} \om{\mechanism{}-sig} PUF is lower than the best \loiss{state-of-the-art DRAM PUF.}} 

\begin{table}[h]
    \caption{Evaluation time of DRAM Latency PUF, PreLatPUF, and \ommmm{\mechanism{}-sig} PUF, using 8KB memory segments.}
    \centering
    \footnotesize{}
    \begin{tabular}{ c  c  c }
      \toprule
        \multirow{2}{*}{\ommm{\bf Latency PUF}} & {\bf PreLatPUF}&  {\bf \ommm{\mechanism{}-sig PUF}}  \\
         & {\bf \ommm{w/(w/o) filter}} & {\bf \ommm{w/(w/o) filter}}  \\
        \midrule
        88.2 ms & 7.95 (1.59) ms & 4.41 (0.88) ms  \\
        \bottomrule
    \end{tabular}
    \label{table:latency_PUFs}
\end{table}

\subsubsection{Randomness Analysis}
\label{sec:nist}

A secure key or seed should have high-entropy. Although we already demonstrated the uniqueness of the responses between different memory segments (Section~\ref{sec:emulating_upla}), this does not guarantee that they \onur{have} high-entropy, and hence that they are suitable to be used as cryptographic keys. 

\vspace{3pt}\noindent\textbf{Methodology.} We analyze the randomness of the values generated by \upla{} with \emph{real} DRAM chips (Table~\ref{table:dram_chips}), with the experimental setup of Section~\ref{sec:emulating_upla}. We use the NIST statistical test suite~\cite{rukhin2001statistical} to analyze the numbers generated by \upla{}.

\vspace{3pt}\noindent\textbf{Results.} We run the NIST test suite with \onur{250KB random streams composed of responses} to different challenges from \onur{all} tested DRAM chips. We use \onur{a} Von Neumann extractor~\cite{shaltiel2011introduction} for whitening the random stream. Our results show that the numbers generated by \upla{} pass all 15 NIST tests, which demonstrates that our PUF is able to generate \om{high-}quality random numbers. \loiss{Appendix~\ref{sec:NIST} has more details.}

\subsection{Preventing Cold Boot Attacks}
\label{sec:eval_coldBootAttacks}

We compare our new \onur{CODIC-based} self-destruction mechanism \onur{(introduced in Section~\ref{sec:coldbootattacks})} to (1)~a mechanism that resets the DRAM's contents by issuing write commands from the memory controller \onur{to every single row in a DRAM chip}, as described in the TCG specification~\cite{computing2009tcg}; (2)~a self-destruction mechanism that uses RowClone~\cite{seshadri2013} to copy rows initialized to zero into \onur{all} other rows \onur{in a DRAM chip}; and (3)~a self-destruction mechanism that uses LISA-clone~\cite{Chang2016LISA} to copy rows initialized to zero into \onur{all} other rows \onur{in a DRAM chip}.

\vspace{3pt}\noindent\textbf{Methodology.} We customize Ramulator\onur{~\cite{kim2016,ramulator.github}} to support \lois{our \mechanism{} implementation}, RowClone and LISA-clone. Table~\ref{table:configuration} shows the summary of the \om{CPU, memory controller, and DRAM parameters} used in our evaluation.

{\setlength{\tabcolsep}{4pt}
\begin{table}[h]
    \footnotesize{}
    \caption{\om{Major Ramulator parameters}.}
    \centering

    \begin{tabular}{ r | l }
        \toprule
        {\bf \om{CPU}} & in-order core, 32KB L1 D\&I, 512KB L2 \\
        {\bf Memory Controller} &  64/64-entry read/write queue,  FR-FCFS~\cite{Rixner:2000,zuravleff1997controller}\\
        {\bf  DRAM} &   \onur{1 channel}, DDR3-1600 x8 11/11/11 \\
        \bottomrule
    \end{tabular}
    \label{table:configuration}
\end{table}
}

Our baseline is the \onur{firmware-based} TCG cold boot attack prevention mechanism. We evaluate TCG by simulating \om{the} firmware that overwrites the memory with zeros by issuing regular write requests. To force writing back the data to memory from cache, we use an instruction that invalidates the data \onur{in} cache (i.e., the \emph{CLFLUSH} instruction in \om{x86~\cite{x86}}). \onur{TCG does not require any hardware changes}.

We implement three self-destruction mechanisms (using \mechanism{}, RowClone, and LISA-clone) that perform back-to-back \mechanism{}/RowClone/LISA-clone commands to each DRAM row (parallelized across banks) while meeting the JEDEC timing specifications (e.g., tRCD, tFAW).
Note that our CODIC self-destruction mechanism could be \om{further} (1)~highly-optimized for performance by DRAM vendors, as they know the exact internal DRAM internal power restrictions, or (2)~highly-optimized for cost, by reusing the self-refresh circuits, as we describe in Section~\ref{sec:self-destruction}.\footnote{We do not evaluate a self-destruction implementation that reuses the self-refresh circuitry because DRAM vendors do not release details about the implementation of self-refresh.} 
We use the \lois{energy and latency of \mechanism{} reported} in Table~\ref{table:energy_latency} (Section~\ref{sec:substrate})\om{. We} calculate the energy \onur{to destroy \omm{contents of the entire DRAM}} using a customized version of DRAMPower~\cite{chandrasekar2012drampower}\om{. We} calculate the latency to destroy the entire content of DRAM using Ramulator.

\begin{figure}[h] \centering
    \includegraphics[width=0.99\linewidth]{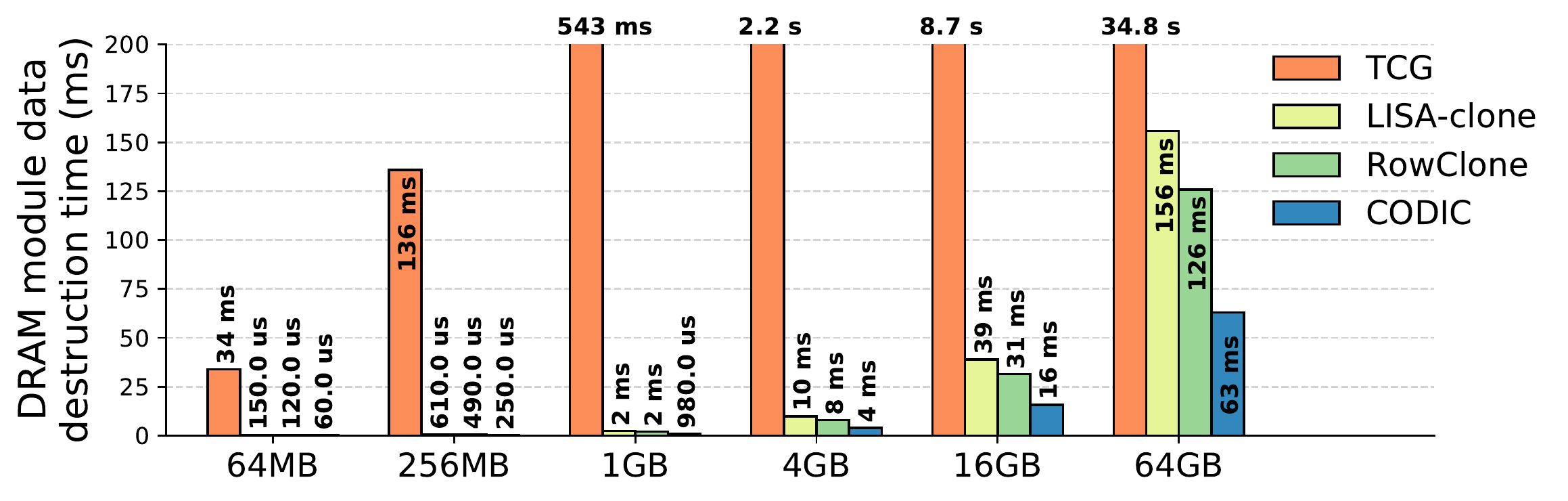}
        \vspace{-0.2cm}
    \caption{Time to destroy all DRAM data in a \onur{DRAM module}.}
    \label{fig:time_destroy}
\end{figure}

\vspace{3pt}\noindent\textbf{Latency Results.}
Figure~\ref{fig:time_destroy} shows the destruction time of a TCG software implementation, \om{and three self-destruction mechanisms that \omm{respectively} use LISA-clone, RowClone, and \mechanism{} to destroy all data in a DRAM module.} 

We test different DRAM \om{module} sizes, from 64MB, used in memories specifically designed for low cost devices~\cite{HyperRAM}, to a hypothetical single-rank 64GB.\footnote{\onur{Existing 64GB DRAM modules today have 4 ranks of 16GB \om{each}~\cite{64GB} to avoid excessive refresh overheads~\cite{Liu2013}}.} Our simulator takes into account all timing parameters defined by the DDRx standard~\cite{jedec2012jedec}. The timing parameters for each size are taken from public datasheets released by vendors~\cite{micronddr3}. For the memories that we \om{do not} have enough information about timing parameters (e.g., 64MB, 64GB), we extrapolate the parameters from existing memory modules.

We make two major observations. 
First, self-destruction using \mechanism{} performs \om{552.7$\times$/\om{2.5}$\times$/\om{2.0}$\times$} faster than TCG/LISA-clone/RowClone \om{for an 8 GB DRAM \ommm{module.}}
Second, the TCG \onur{provides lower} latency \onur{for} small memory sizes (e.g., 64MB, 256MB), but the latency for 1GB and larger sizes is \onur{prohibitively large}. 

We conclude that self-destruction using CODIC destroys the entire content of DRAM significantly faster than the best state-of-the-art mechanisms.

\vspace{3pt}\noindent\textbf{Energy Results.} Our results show that self-destruction using CODIC consumes \om{41.7$\times$}\om{/2.5$\times$/1.7$\times$} less energy than TCG/LISA-clone/RowClone, \om{for an 8 GB DRAM module.}

\vspace{3pt}\noindent\textbf{Comparison with Other State-of-the-Art Mechanisms.}
There exist other mechanisms that protect against cold boot attacks that are fundamentally different \onur{from} our approach. This is the case with memory encryption, which provides strong security guarantees at the cost of additional energy consumption \onur{and complexity}. Table~\ref{table:ColdBoot_comparison} shows the performance, power, and area overhead of our self-destruction mechanism compared to ChaCha-8\onur{~\cite{bernstein2008chacha,Yitbarek2017}} and AES-128\onur{~\cite{daemen2001reijndael,Yitbarek2017}}, two low-cost ciphers that can be used to prevent cold boot attacks efficiently~\cite{Yitbarek2017}.

\begin{table}[h]
    \caption{Overhead of \om{\mechanism} self-destruction \omm{vs.} two other mechanisms to prevent cold boot attacks on an Intel Atom N280 processor.}
    \vspace{-0.1cm}
    \centering
    \begin{threeparttable}
    \centering
    \scriptsize{}
    \setlength\tabcolsep{2pt} 
    \setlength\extrarowheight{2pt}
    \begin{tabular}{ r  c  c  c}
      \toprule
        \multicolumn{1}{c}{} & \multicolumn{1}{c}{\bf \mechanism{} Self-Dest.}  & {\bf ChaCha-8 } & {\bf AES-128 }  \\
         \cmidrule(lr){2-2} \cmidrule(lr){3-3} \cmidrule(lr){4-4}
        {\bf Runtime Performance \om{Overhead}} & \omm{$\sim$}0\% & \omm{$\sim$}0\% & \omm{$\sim$}0\%\tnote{1}   \\
        {\bf Runtime Power \om{Overhead}\tnote{2}~} & \omm{$\sim$}0\% & \omm{$\sim$}17\% & \omm{$\sim$}12\%   \\
        {\bf \onur{Area Overhead (Processor/DRAM)}} & \onur{\omm{$\sim$(0.0/1.1)}\%} & \onur{\omm{$\sim$(0.9/0.0)}\%} & \onur{\omm{$\sim$(1.3/0.0)}\%}  \\
        \bottomrule
    \end{tabular}
    \begin{tablenotes}
     \item[1] \onur{Assuming that the maximum number of} back-to-back row hits \onur{is 16}.
     \item[2] At peak \om{memory} bandwidth utilization.
    \end{tablenotes}
    \end{threeparttable}
    \label{table:ColdBoot_comparison}
\end{table}

We make \onur{three} main observations. First, our self-destruction mechanism has \emph{zero} performance and power overhead at runtime.
Second, although ChaCha-8 and AES-8 can be implemented for hiding the encryption latency in the common case~\cite{Yitbarek2017}, the power overhead is \onur{very} significant \onur{(17\% for ChaCha-8 and 12\% for AES-128)} in low-cost processors such as the Intel Atom \om{N280~\cite{N280}}. \onur{Third, the processor area overhead \om{is low} for ChaCha-8 (0.9\%) and AES-128 (1.3\%), whereas \mechanism{} does not require processor changes, and the DRAM area overhead \om{is low} for \mechanism{} (1.1\%), whereas ChaCha-8 and AES-128 do not require DRAM changes}. We conclude that our zero-runtime-overhead proposal is a very \om{power- and area-} efficient way to protect against cold boot attacks in systems where encryption is \om{expensive or otherwise undesirable}.\footnote{While AES-128 and ChaCha-8 provide additional security features, we evaluate their ability to prevent cold boot attacks, as studied in recent literature~\cite{Yitbarek2017}.}

%% file: 08_related.tex
\section{Related Work}
\label{sec:relatedWork}

To our knowledge, this is the first paper to propose a low-cost substrate to control DRAM internal circuit timings\om{, thereby enabling} \onur{the implementation of} \om{a variety of} new features\onur{, and custom} \ommmm{reliability,} performance \onur{and energy} optimizations in DRAM. We demonstrate \mechanism{}'s flexibility with two \mechanism{} \ommmm{configurations that enable the implementation of two security applications that improve the} state-of-the-art: (1)~PUF-based authentication, and (2)~cold boot attack prevention.

\vspace{3pt}\noindent\textbf{PUFs.}
%
PUFs have been proposed using different substrates (e.g., SRAM~\cite{holcomb2007initial,holcomb2009power, Bhargava2012, zheng2013, xiao2014, Bacha2015}, ASIC logic~\cite{Daihyun2005, vanderLeest:2010}, DRAM~\cite{Tehranipoor:2015, Sutar2016, Xiong2016, kim2018, talukder2019prelatpuf}, \onur{or FPGA~\cite{Gassend:2002}})\om{. We} focus our comparison against DRAM-based PUFs as DRAM has a large address space and is a \om{widely-used} technology in systems today.
Section~\ref{sec:physicallyUnclonable} already \om{extensively compares the CODIC-based PUF} against two state-of-the-art DRAM-based PUFs: the DRAM Latency PUF~\cite{kim2018} and PreLatPUF~\cite{talukder2019prelatpuf}.
We show that our \mechanism-based PUF provides \lois{(1)~robust responses, (2)~lower evaluation time, and (3)~resiliency to temperature changes,} characteristics that other DRAM PUFs do not  provide all together.

\vspace{3pt}\noindent\textbf{Cold Boot Attacks.}
We have already discussed several cold boot attack prevention mechanisms~\cite{Suh:2003, arnold2004ibm, Yang:2005, Duc2006, Rogers2007, Henson:2014, amd_sme, Yitbarek2017, costan2016intel}, and compared against\onur{~\cite{computing2009tcg,Yitbarek2017}} in Section~\ref{sec:coldbootattacks}.
A previous work on data lifetime management~\cite{lee2017dram} proposes to disable access of untrusted programs to data in DRAM using a new flag in the DRAM decoder, controlled by a DRAM command. However, this mechanism does not prevent an attacker \onur{who has} \emph{physical access} to the device from reading data freely.
Seol et al.~\cite{Seol2017} propose a mechanism to initialize DRAM with a reset operation based on connecting/disconnecting power lines. This reset operation has a larger latency than \mechanism{}, as it requires a precharge and an \om{activate command} \onur{to each row}, while \mechanism{} requires only one command to destroy the contents of a \onur{row}. 
Our mechanism for protecting against cold boot attacks improves upon the state-of-the-art by proposing a very simple mechanism with no performance or energy overhead at runtime \onur{(Section~\ref{sec:coldbootattacks})}. 

\onur{Amnesiac DRAM~\cite{seol2019amnesiac}, published after our preprint version of \mechanism{}~\cite{orosa2019dataplant}, proposes a mechanism \om{similar} to our method for protecting against cold boot attacks (Section~\ref{sec:coldbootattacks}). \om{Our} cold boot attack prevention mechanism is \om{only one} use case \om{of} \mechanism{}, which can enable many other applications, as we discuss in Section~\ref{sec:otheraplications}.}

\vspace{3pt}\noindent\textbf{Processing-in-Memory (PIM) Security Mechanisms.} Previous works propose to obfuscate memory access patterns in (1)~3D/2.5D memories with logic capabilities~\cite{Awad:2017,Aga:2017}, and (2)~commodity DRAM \om{modules} with an additional secure chip~\cite{shafiee2018secure}. These approaches are orthogonal \om{to} \mechanism{}. \onur{Thus,} they can be \onur{combined with \mechanism{}} to enhance the overall security capabilities of main memory, \onur{especially PIM systems}.

%% file: 09_conclusion.tex
\section{Conclusion}

We propose \mechanism{}, a low-cost DRAM substrate that enables fine-grained control over DRAM internal circuit timings. \mechanism{} can be used to enable important existing and novel mechanisms and DRAM optimization techniques at low-cost on any device that uses DRAM. To demonstrate the potential of \mechanism{}, we implement (1)~a new Physical Unclonable Function (PUF), and (2)~a new cold boot attack prevention mechanism. We perform a thorough evaluation that shows that our two new mechanisms perform better than the best state-of-the-art mechanisms. \onur{We conclude that the \mechanism{} can be used for implementing very efficient security applications at low cost. We hope and believe that \mechanism{} will inspire and enable (1)~\om{other} new DRAM functionalities that enhance the applications \om{and functionalities} of DRAM, and (2)~new DRAM reliability, performance and energy optimizations.}

%% file: 10_secureDeallocation.tex
\section{Secure Deallocation}
\label{sec:secureDeallocationEval}
\label{sec:eval_dealloc}

\label{sec:othermechanisms}

\lois{In this Section, we describe and evaluate secure deallocation, a security application that can be efficiently implemented with any \mechanism{} primitive (see Section~\ref{sec:substrate:implementing}). We note that since most Operating Systems require newly allocated memory to be filled with \loiss{zero \loisss{values~\cite{bovet2005understanding,chow2005shredding,windowsInternals,Yang2011}}.} \dtran{} (Section~\ref{sec:dtran}) is the best candidate for secure deallocation, since \dtran{} efficiently zeroes memory during deallocation.}

\loiss{Modern} applications (e.g., web servers, web browsers, word processors) do not immediately \lois{erase} data from memory when it is no longer needed. Instead, \loiss{the Operating System (OS) has the responsibility of physically erasing the data \loiss{in a memory region} when the program deallocates \loiss{the memory region}. However,  the OS \loisss{may erase} deallocated data} \loisss{just before} the \loiss{physical} memory is required for \loiss{allocating} new data. As a consequence, \emph{sensitive data could remain in memory for an indefinite amount of time}, which increases the risk of \loiss{data leakage}.

Secure deallocation~\cite{chow2005shredding, Anikeev:2013, Sha2018, harrison2007protecting, Garfinkel:2004} is a technique that resets data in memory to zero at the moment of \loiss{deallocation}. This technique reduces the time that critical data is vulnerable to attacks \loiss{that scan through memory to recover sensitive data}. 
\mechanism{} enables the implementation of \loisss{secure deallocation} techniques with very low latency, energy, and area overhead. 

\vspace{3pt}\noindent\textbf{Evaluation Methodology.} We compare LISA-clone~\cite{Chang2016LISA}, RowClone~\cite{seshadri2013} and  \dtran{} (Section~\ref{sec:dtran}) to a software secure deallocation mechanism~\cite{chow2005shredding} that fills memory with zero values using common memory write instructions. We \lois{customize} Ramulator~\cite{kim2016} to support all evaluated mechanisms on in-order cores. To generate the traces that drive our simulator, we use Pin~\cite{Luk:2005} for user-level traces, and \lois{the Bochs~\cite{bochs}} full-system emulator to generate the memory traces that include Linux kernel page allocations and deallocations. 
To estimate the energy \lois{consumption of the DRAM module}, we use a customized version of DRAMPower~\cite{chandrasekar2012drampower}. Table~\ref{table:configuration_global} shows the system configuration used \lois{in} our evaluation.

{\setlength{\tabcolsep}{4pt}
\begin{table}[h]
    \centering
    \footnotesize{}
    \caption{System configuration.}\label{table:configuration_global}
    \begin{tabular}{ r | l }
        \toprule
        {\bf Processor} &  1-4 cores, in-order, \\ [0.5ex]
        {\bf Cache} &  L1:64KB, L2:512KB per core, 64B lines\\
        {\bf Memory Controller} &  64/64-entry read/write queue,  FR-FCFS~\cite{Rixner:2000,zuravleff1997controller}\\
        {\bf  DRAM} &   1 channel, DDR3-1600 x8 11/11/11 \\
        \bottomrule
    \end{tabular}
    \vspace{1.5mm}
    
\end{table}
}

\lois{Table~\ref{table:benchmarks_init} describes the \lois{6} memory-allocation-intensive benchmarks that we use. For the multicore evaluation (4 cores), we choose 50 mixes of workloads, \loiss{where} each mix is composed of two memory-allocation-intensive benchmarks and two non-memory-allocation-intensive benchmarks. The non-memory-allocation-intensive benchmarks are TPC-C~\cite{tpc}, \loiss{TPC-H~\cite{tpc},} STREAM~\cite{stream}, SPEC2006~\cite{spec2006}, DynoGraph (pagerank, bfs, stream)~\cite{dynograph}, and HPCC RandomAccess~\cite{hpc}. Table~\ref{table:multicore_benchmarks} shows 5 representative benchmark mixes.}

\begin{table}[h]
    \centering
    \caption{\lois{Memory-allocation-intensive benchmarks used} for evaluating secure deallocation.}
    \footnotesize{}
    \begin{tabular}{ r | l }
      \toprule
        {\bf\emph Benchmark} & {\bf\emph Description} \\
        \midrule
        mysql &  MySQL~\cite{mysql} loading the sample {\it employeedb}.\\
        mcached &  Memcached~\cite{memcached}, a memory object caching system \\
        compiler &  Compilation phase \loiss{of} the GNU C compiler \\
        bootup &  Linux kernel \loiss{boot-up} phase \\
        shell &  Script running 'find' in a directory tree with 'ls' \\
        malloc & stress-ng~\cite{stress-ng} stressing the malloc primitive \\
     \bottomrule
    \end{tabular}
    \label{table:benchmarks_init}
\end{table}
\vspace{3mm}
\begin{table}[h]
    \centering
    \caption{Five representative mixes (out of 50) used in the multicore evaluation for secure deallocation.}
    \scriptsize{}
    \setlength\tabcolsep{1.5pt}
    \begin{tabular}{ r  l r l}
      \toprule
        {\bf MIX1:} & malloc, bootup, tpcc64, libquantum & {\bf MIX4:} & malloc, shell, xalancbmk, bzip2 \\
        \midrule
        {\bf MIX2:} & shell, bootup, lbm, xalancbmk & {\bf MIX5:} &malloc, malloc, astar, condmat\\ %
        \midrule
        {\bf MIX3:} & bootup, shell, pagerank, pagerank &&\\
        \bottomrule
    \end{tabular}
    \label{table:multicore_benchmarks}
\end{table}

\begin{figure}[h] \centering
    \includegraphics[width=1.0\linewidth]{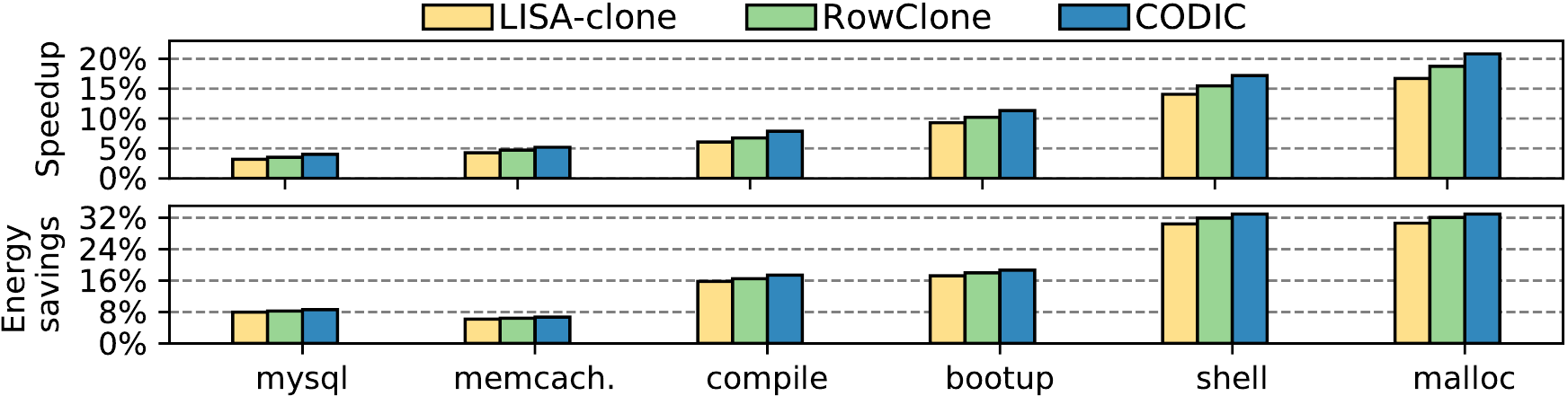}
    \caption{Single core speedup (larger is better) and energy savings (larger is better) of the secure deallocation hardware approaches \loiss{over} a software approach.}
    \label{fig:single_core}
\end{figure}

\vspace{3pt}\noindent\textbf{Evaluation Results.}
Figure~\ref{fig:single_core} shows the \lois{single-core speedup} (higher is better) and energy savings (\lois{higher} is better) of LISA-clone, RowClone and  \mechanism{} normalized to a software secure deallocation implementation that fills memory with zero values using common memory write instructions. 

\lois{We make two observations}. First, \loiss{hardware approaches (LISA-clone, RowClone, \mechanism{})} improve \loiss{performance} by up to 21\% and \loiss{energy} savings by up to 34\%, compared to the \loisss{baseline} software \loiss{approach}. Second, \mechanism{} performs better \loiss{and consumes less energy} than LISA-clone and RowClone \loiss{for all workloads.} 

\begin{figure}[h] \centering
    \includegraphics[width=1.0\linewidth]{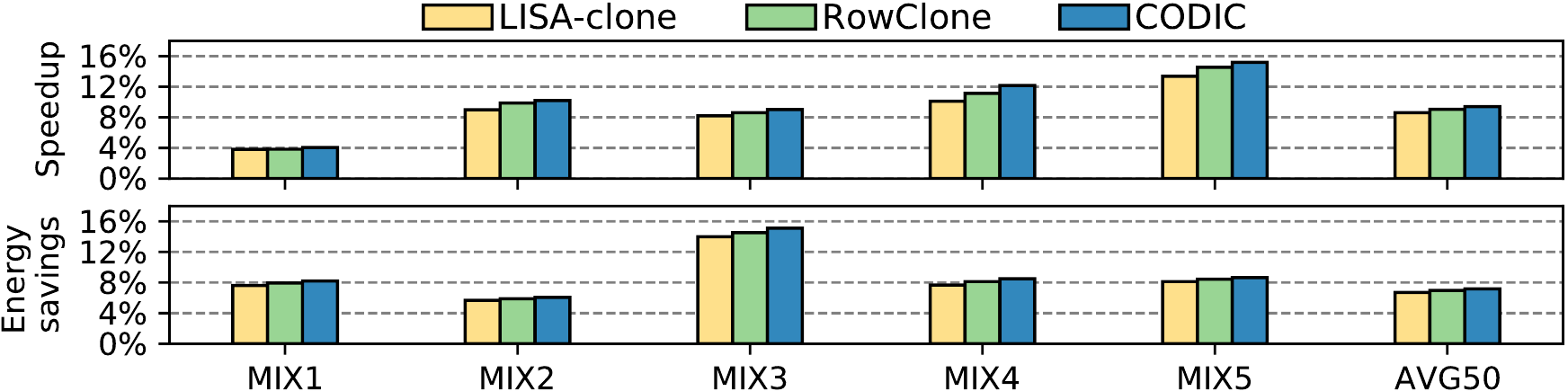}
    \caption{4-core speedup (larger is better) and energy savings (larger is better) of the secure deallocation hardware approaches \loiss{over} a software approach.}
    \label{fig:multi_core}
\end{figure}

\lois{Figure~\ref{fig:multi_core} shows the \lois{speedup} and energy savings of \mechanism{} and other state-of-the-art mechanisms in a 4-core processor, normalized to the performance and energy consumption of the software secure deallocation implementation}. 

We make the same two observations \lois{as} for a single core processor: 1) \loiss{hardware} approaches \loiss{perform better than} the software \loiss{approach}, and 2) \mechanism{}  performs better than LISA-clone and RowClone.

We have demonstrated the flexibility of the \mechanism{} substrate with our secure deallocation mechanism, in addition to a better DRAM PUF (Section~\ref{sec:physicallyUnclonable}) and an efficient cold boot attack prevention mechanism (Section~\ref{sec:coldbootattacks}). Our evaluation shows that our \mechanism{}-based secure deallocation mechanism has higher performance and energy efficiency compared to the state-of-the-art. We conclude that in addition to being flexible, \mechanism{} is a very powerful substrate that enables \lois{more secure and efficient} \loiss{security} mechanisms. 

%% file: 12_NIST.tex
\section{NIST Test Suite \loiss{Results}}
\label{sec:NIST}

 Table~\ref{table:NIST} shows the average NIST~\cite{rukhin2001statistical} p-values and NIST final results for signatures generated by \upla{} from different challenges. 

\begin{table}[h]
    \caption{\upla{} average results of the NIST statistical test suite. }
    \centering
    \footnotesize{}
    \setlength\tabcolsep{5pt} 
    \begin{tabular}{rcc}
       \toprule
        {\emph NIST Test} &  \loiss{\emph p-value} & {\emph Result} \\
        \midrule
        monobit & 0.681 & PASS\\
        frequency\_within\_block & 1.000 & PASS\\
        runs & 0.298 & PASS\\
        longest\_run\_ones\_in\_a\_block & 0.287 & PASS\\
        binary\_matrix\_rank & 0.536 & PASS\\
        dft & 0.165 & PASS\\
        non\_overlapping\_template\_matching & 0.808 & PASS\\
        overlapping\_template\_matching & \lois{0.210} & \lois{PASS}\\
        maurers\_universal & 0.987 & PASS\\
        linear\_complexity & 0.0185 & PASS\\
        serial & 0.988 & PASS\\
        approximate\_entropy & 0.194 & PASS\\
        cumulative\_sums & 0.940 & PASS\\
        random\_excursion & 0.951 & PASS\\
        random\_excursion\_variant & 0.693 & PASS\\
        \bottomrule
    \end{tabular}
    \label{table:NIST}
\end{table}

Our results show that the \loisss{values} generated by \upla{} pass all 15 NIST \loisss{randomness} tests, which demonstrates that our PUF is able to return \om{high-}quality random signatures \loiss{for} different challenges.

%% file: 13_newCODIC_variant.tex
\section{\uesa{}}
\label{sec:CODIC_new_variant}

This \loiss{section} describes \uesa{}, an alternative \mechanism{} variant for generating signature values.
\uesa{} generates signature values by sensing and amplifying a precharged bitline that is \loiss{\emph{not}} connected to the DRAM cell (i.e., the wordline is not \loiss{raised}).
By doing so, the SA amplifies \lois{the bitline voltage} to a signature value that 1) \loiss{does not} depend on the charge \loiss{level} of \loiss{any} cell, and 2) depends \loiss{only} on the SA process variation \loiss{present in the SA and the bitline}. \lois{Once the SA drives the bitline towards the final restored value, \uesa{} can optionally write this value into \loiss{a} cell} by \lois{raising the wordline}.

\begin{figure}[h] \centering
    \includegraphics[width=0.5\linewidth]{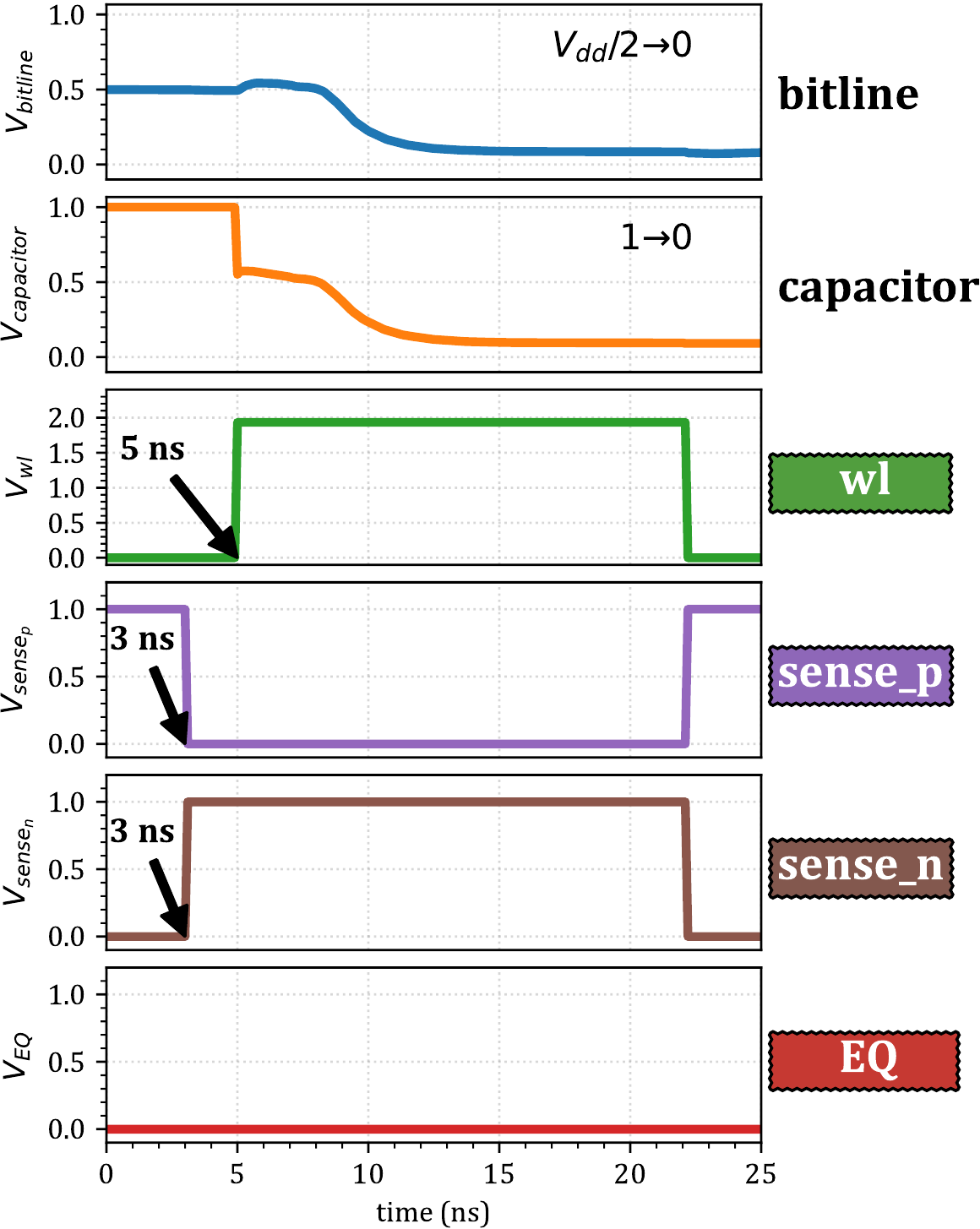}%
    \caption{\uesa{}.}
    \label{fig:new_variant}
\end{figure}

Figure~\ref{fig:new_variant} shows the SPICE simulation of \uesa{}. \uesa{} raises the \emph{sense\_p} and \emph{sense\_n} signals (at 3ns) before it raises the \emph{wl} signal (at 5ns).  Because the bitline is initially precharged to $V_{dd}/2$ voltage, the SA (\emph{sense\_n} and \emph{sense\_p}) amplifies \loiss{the sensed value of the bitline} to zero or one depending \loiss{purely} on process variation (in this example, the SA amplifies \loiss{the bitline value} to zero).

\vspace{3pt}\noindent\textbf{Evaluation Methodology.} We evaluate \uesa{} as a PUF with SPICE simulations. \lois{Unfortunately,} it is \loiss{infeasible} to conduct experiments on real DRAM chips, as \uesa{} requires \minp{changes to the internal DRAM timings, which are hard-wired in commodity DRAM chips}. To show the effects of process variation on the values generated by \uesa{}, we evaluate a detailed SA SPICE model \lois{(see Section~\ref{sec:eval_lat-ener-area})} using Monte Carlo simulations. 
We model variations in all the affected components of the \lois{SAs} (transistor length/width/threshold voltage). Our SA model always generates \loiss{`1' values} in absence of process variation. When we introduce process variation into the simulation, we observe that some SAs generate \loiss{`0' values} as well (i.e., some SAs generate bit flips). We run 100,000 simulations for each variation.

\vspace{3pt}\noindent\textbf{Evaluation Results.}
Table~\ref{table:common_design} shows the percentage of SAs that generate bit flips with \uesa{}. The positions of the SAs that cause bit flips with \uesa{} are random, and depend on process variation. We perform simulations for different levels of process variation and temperature.

\begin{table}[h]
    \caption{Effect of \lois{Process Variation (PV)} and temperature on the percentage of \uesa{} bit flips.}
    \centering
    \footnotesize{}
    \setlength\tabcolsep{2pt} 
    \begin{tabular}{c | cccc |cccc | }
        \toprule
        \multicolumn{1}{c|}{ } & \multicolumn{4}{c|}{ {\emph PV effects}} & \multicolumn{4}{c|}{{\emph Temperature effects \lois{(4\% PV)}}} \\
         & {\bf 2\%} & {\bf 3\%} & {\bf 4\%} & {\bf 5\%} &  {\bf 30$^\circ$C} & {\bf 60$^\circ$C} & {\bf 70$^\circ$C} & {\bf 85$^\circ$C}  \\
        \cmidrule(lr){2-5} \cmidrule(lr){6-9}
          \multicolumn{1}{c|}{\bf Bit flips} & 0.00\% & 0.00\% & 0.02\% & 0.19\%  & 0.02\% & 0.19\% & 0.21\%  & 0.15\% \\
        \bottomrule
    \end{tabular}
    \label{table:common_design}
\end{table}

We make three main observations. First, small process \loisss{variation amounts} ($<$4\%) are not enough to generate \uesa{} bit flips. Second, large process variation \loisss{amounts} increase the number of SAs that generate bit flips with \uesa{}. As the technology scales, process variation becomes more significant, which increases the number of bit flips generated by \uesa{}.  Third, temperature changes do not cause significant variations in the number of SAs that cause bit flips with \uesa{}. 

%% file: 11_DRAMchips.tex
\section{\loiss{Evaluated} DRAM Chips}
\label{sec:DRAMchips}

\begin{table}[h]
    \caption{Characteristics of the 15 evaluated \om{DDR3} DRAM modules (136 DDR3 DRAM chips).}
    \centering
    \scriptsize{}
    \setlength\tabcolsep{4pt} 
    \begin{tabular}{rccccccl}
        \toprule
        {\emph Module} & {\emph Vendor} &  {\emph Chips}  & {\emph Pins} & {\emph Ranks} & \makecell{\emph Chip size\\(Gb)} & \makecell{\emph Freq.\\(MT/s)} & \multicolumn{1}{c}{\emph Voltage}  \\
        \midrule
        \stripe
        $M1$ & A &  8 & $\times$8 & 1 & \om{4} &  1600  & 1.35V (DDR3L) \\ 
        $M2$ & A  & 8 & $\times$8 & 1 & \om{4} & 1600  & 1.35V (DDR3L) \\ 
        \stripe
        $M3$ & A  & 8 & $\times$8 & 1 & \om{4} & 1600  & 1.35V (DDR3L) \\ 
        $M4$ & A & 8 & $\times$8 & 1 & \om{4} & 1600 & 1.35V (DDR3L) \\ 
        \stripe
        $M5$ & A & 8 & $\times$8 & 1 & \om{4} & 1600 & 1.50V (DDR3) \\ 
        $M6$ & A & 8 & $\times$8 & 1 & \om{4} & 1600 & 1.50V (DDR3) \\ 
        \stripe
        $M7$ & A & 8 & $\times$8 & 1 & \om{4} & 1600 & 1.50V (DDR3) \\ 
        $M8$ & A & 8 & $\times$8 & 1 & \om{4} & 1600 & 1.50V (DDR3) \\ 
        \stripe
        $M9$ & B  & 16 & $\times$8 & 2 & \om{2} & 1333 & 1.50V (DDR3) \\ 
        $M10$ & B  & 16 & $\times$8 & 2 & \om{2} & 1333 & 1.50V (DDR3) \\ 
        \stripe
        $M11$ & B & 8 & $\times$8 & 1 & \om{4} & 1600 & 1.35V (DDR3L) \\ 
        $M12$ & C & 8 & $\times$8 & 1 & \om{4} & 1600 & 1.35V (DDR3L) \\ 
        \stripe
        $M13$ & C & 8 & $\times$8 & 1 & \om{4} & 1600 & 1.35V (DDR3L) \\ 
        $M14$ & C & 8 & $\times$8 & 1 & \om{4} & 1600 & 1.35V (DDR3L) \\ 
        \stripe
        $M15$ & C & 8 & $\times$8 & 1 & \om{4} & 1600 & 1.35V (DDR3L) \\ 
        \bottomrule
    \end{tabular}
    \label{table:dram_chips}
\end{table}